\documentclass[journal]{IEEEtran}
\usepackage{amsmath,soul}
\usepackage{graphicx}

\usepackage{amsfonts}
\usepackage{mathabx}
\usepackage{color}
\usepackage{epstopdf} 

\newcommand{\smallsub}[1]{\! {\scriptscriptstyle \mathcal{#1}} }

\newtheorem{proposition}{Proposition}

\newtheorem{definition}{Definition}
\newtheorem{remark}{Remark}
\newtheorem{lemma}{Lemma}

\newtheorem{example}{Example}

\newtheorem{theorem}{Theorem}
\newtheorem{property}{Property}
\newtheorem{condition}{Condition}

\newcommand{\tn}{\hat\theta_N}

\newcommand{\Eb}{\bar{\mathbb{E}}}
\newcommand{\M}{\mathcal{M}}
\newcommand{\E}{\mathbb{E}}

\newcommand{\mY}{\mathcal{Y}}
\newcommand{\mD}{\mathcal{D}}
\newcommand{\mZ}{\mathcal{Z}}
\newcommand{\mQ}{\mathcal{Q}}
\newcommand{\mA}{\mathcal{A}}
\newcommand{\mB}{\mathcal{B}}
\newcommand{\mF}{\mathcal{F}}

\newcommand{\mL}{\mathcal{L}}
\newcommand{\mU}{\mathcal{U}}
\newcommand{\mX}{\mathcal{X}}
\newcommand{\smU}{\smallsub{U}}
\newcommand{\smZ}{\smallsub{Z}}

\newcommand{\smA}{\smallsub{A}}
\newcommand{\smB}{\smallsub{B}}
\newcommand{\smF}{\smallsub{F}}

\newcommand{\smD}{\smallsub{D}}
\newcommand{\smY}{\smallsub{Y}}
\newcommand{\smQ}{\smallsub{Q}}
\newcommand{\smX}{\smallsub{X}}
\newcommand{\smL}{\smallsub{L}}

\newcommand{\beq}{\begin{equation}}
\newcommand{\eeq}{\end{equation}}
\newcommand{\beqs}{\begin{equation*}}
\newcommand{\eeqs}{\end{equation*}}
\newcommand{\beqr}{\begin{eqnarray}}
\newcommand{\eeqr}{\end{eqnarray}}
\newcommand{\beqrs}{\begin{eqnarray*}}
	\newcommand{\eeqrs}{\end{eqnarray*}}

\usepackage{cite}

\begin{document}
%
\title{A local direct method for module identification in dynamic networks with correlated noise}
%
%
%

\author{Karthik R. Ramaswamy,~\IEEEmembership{Student Member,~IEEE,}
        Paul M.J. Van den Hof,~\IEEEmembership{Fellow,~IEEE}
\thanks{Karthik Ramaswamy and Paul Van den Hof are with the Department of Electrical
	Engineering, Eindhoven University of Technology, Eindhoven, The
	Netherlands {\tt\small \{k.r.ramaswamy, p.m.j.vandenhof\}@tue.nl}}
\thanks{Paper submitted for publication 2 August 2019. Revised: 16 January 2020 and 10 July 2020. Final version 31 October 2020. This work has received funding from the European Research Council (ERC), Advanced Research Grant SYSDYNET, under the European Union's Horizon 2020 research and innovation programme (grant agreement No 694504).}
}

%



\maketitle

\begin{abstract}
The identification of local modules in dynamic networks with known topology has recently been addressed by formulating conditions for arriving at consistent estimates of the module dynamics, under the assumption of having disturbances that are uncorrelated over the different nodes. The conditions typically reflect the selection of a set of node signals that are taken as predictor inputs in a MISO identification setup. In this paper an extension is made to arrive at an identification setup for the situation that process noises on the different node signals can be correlated with each other. In this situation the local module may need to be embedded in a MIMO identification setup for arriving at a consistent estimate with maximum likelihood properties. This requires the proper treatment of confounding variables. The result is a set of algorithms that, based on the given network topology and disturbance correlation structure, selects an appropriate set of node signals as predictor inputs and outputs in a MISO or MIMO identification setup.
Three algorithms are presented that differ in their approach of selecting measured node signals. Either a maximum or a minimum number of measured node signals can be considered, as well as a preselected set of measured nodes.
\end{abstract}


%
\IEEEpeerreviewmaketitle

\section{Introduction}\label{sec-intro}
%
%
%
%
\IEEEPARstart{I}{n} recent years increasing attention has been given to the development of new tools for the identification of large-scale interconnected systems, also known as dynamic networks. These networks are typically thought of as a set of measurable signals (the node signals) interconnected through linear dynamic systems (the modules), possibly driven by external excitations (the reference signals). Among the literature on this topic, we can distinguish three main categories of research. The first one focuses on identifying the topology of the dynamic network \cite{MaterassiSalapaka2012,Sanandaji2011,Materassi10,Chiuso&Pillonetto_Autom:12, Shi&etal_ECC:19}. The second category concerns  identification of the full network dynamics \cite{Haber&Verhaegen:TAC:14,Torres14,Weerts&etal_Autom:18_reducedrank,Weerts&etal_CDC:16,Zorzi&Chiuso:17,Bazanella&etal_CDC:17}, including aspects of identifiability, particularly addressed in \cite{Goncalves&Warnick:08, Weerts&etal_Autom:18_identifiability, Hendrickx&Gevers&Bazanella_TAC:19},
while the third one deals with identification of a specific component (module) of the network, assuming that the network topology is known (the so called local module identification), see \cite{VandenHof&etal_Autom:13,Dankers&etal_Autom:15,Linder&Enqvist_ijc:17,Ramaswamy&etal_CDC:18, Everitt&Bottegal&Hjalmarsson_Autom:18,Gevers&etal:sysid18}.

In this paper we will further expand the work on the local module identification problem. In \cite{VandenHof&etal_Autom:13}, the classical \emph{direct-method} \cite{Ljung:99} for closed-loop identification has been generalized to a dynamic network framework using a MISO identification setup. Consistent estimates of the target module can be obtained when the network topology is known and all the node signals in the MISO identification setup are measured. The work has been extended in \cite{Materassi&Salapaka:CDC15,Dankers&etal_TAC:16,Materassi&Salapaka:20} towards the situation where some node signals might be non-measurable, leading to an additional predictor input selection problem. A similar setup has also been studied in \cite{Ramaswamy&etal_CDC:18}, where an approach has been presented based on empirical Bayesian methods to reduce the variance of the target module estimates. In \cite{Dankers&etal_Autom:15} and \cite{Everitt&Bottegal&Hjalmarsson_Autom:18}, dynamic networks having node measurements corrupted by sensor noise have been studied, and informative experiments for consistent local module estimates have been addressed in \cite{Gevers&etal:sysid18}.

A standing assumption in the aforementioned works \cite{VandenHof&etal_Autom:13}, \cite{Ramaswamy&etal_CDC:18}, \cite{Gevers&etal:sysid18}, \cite{Dankers&etal_TAC:16} is that the process noises entering the nodes of the dynamic network are uncorrelated with each other. This assumption facilitates the analysis and the development of methods for local module identification, reaching
\emph{consistent} module estimates using the direct method. However, when process noises are correlated over the nodes, the consistency results for the considered MISO direct method collapse. In this situation it is necessary to consider also the \emph{noise topology or disturbance correlation structure}, when selecting an appropriate identification setup.  Even though the indirect and two-stage methods in \cite{Dankers&etal_Autom:15,Gevers&etal:sysid18} can handle the situation of correlated noise and deliver consistent estimates, the obtained estimates will not have minimum variance.

In this paper we particularly consider the situation of having dynamic networks with disturbance signals on different nodes that possibly are correlated, while our target moves from consistency only, to also minimum variance (or Maximum Likelihood (ML)) properties of the obtained local module estimates. We will assume that the topology of the network is known, as well as the (Boolean) correlation structure of the noise disturbances, i.e. the zero-elements in the spectral density matrix of the noise. While one could use techniques for full network identification (e.g., \cite{Weerts&etal_Autom:18_reducedrank}), our aim is to develop a method that uses only local information. In this way, we avoid (i) the need to collect node measurements that are ``far away'' from the target module, and (ii) the need to identify unnecessary modules that would come with the price of higher variance in the estimates.

Using the reasoning first introduced in \cite{VandenHof&etal_CDC:17}, we build a constructive procedure that, choosing a limited number of predictor inputs and predicted outputs, builds an identification setup that guarantees maximum likelihood (ML) properties (and thus asymptotic minimum variance) when applying a direct prediction error identification method. In this situation we have to deal with so-called \emph{confounding variables} (see e.g. \cite{VandenHof&etal_CDC:17}, \cite{Dankers&etal_IFAC:17}), that is, unmeasured variables that directly or indirectly influence both the predicted output and the predictor inputs, and lead to lack of consistency.
The effect of confounding variables will be mitigated by extending the number of predictor inputs and/or predicted outputs in the identification setup, thus including more measured node signals in the identification. Preliminary results for the particular ``full input'' case have been presented in \cite{VandenHof&etal_CDC:19}. Here we generalize that reasoning to different node selection schemes, and provide a generally applicable theory that is independent of the particular node selection scheme selected.

This paper is organized as follows. In section \ref{sec:setup}, the dynamic network setup is defined. Section \ref{sec:avail} provides a summary of available results from the existing literature of local module identification related to the context of this paper. Next, important concepts and notations used in this paper are defined in Section \ref{sec:concepts} while the MIMO identification setup and main results are presented in subsequent sections. Sections \ref{sec:fullinput}-\ref{sec:usersel} provide algorithms and illustrative examples for three different ways of selecting input and output node signals: the full input case, the minimum input case, and the user selection case. This is followed by Conclusions. The technical proofs of all results are collected in the Appendix.

\section{Network and identification setup}
\label{sec:setup}

Following the basic setup of \cite{VandenHof&etal_Autom:13}, a dynamic network is built up out of $L$ scalar \emph{internal variables} or \emph{nodes} $w_j$, $j
= 1, \ldots, L$, and $K$ \emph{external variables} $r_k$, $k=1,\ldots K$.
Each internal variable is described as:
\begin{align}
w_j(t) = \sum_{\stackrel{l=1}{l\neq j}}^L
G_{jl}(q)w_l(t) + u_j(t) + v_j(t)
\label{eq:netw_def}
\end{align}
where $q^{-1}$ is the delay operator, i.e. $q^{-1}w_j(t) = w_j(t-1)$,
\begin{itemize}
	\item $G_{jl}$ are proper rational transfer functions, referred to as {\it modules};
	\item There are no self-loops in the network, i.e. nodes are
	not directly connected to themselves $G_{jj} = 0$;
	\item $u_j(t)$ is generated by the \emph{external variables} $r_k(t)$ that can directly be manipulated by the user and is given by
	$u_j(t) = \sum_{k=1}^{K}R_{jk}r_k(t)$ where $R_{jk}$ are stable, proper rational transfer functions;
	\item $v_j$ is \emph{process noise}, where the vector process $v=[v_1 \cdots v_L]^T$ is modelled as a stationary stochastic process with rational spectral density $\Phi_v(\omega)$, such that there exists a white noise process $e:= [e_1 \cdots e_L]^T$, with covariance matrix $\Lambda>0$ such that
	$ v(t) = H(q)e(t)$,
	where $H(q)$ is square, stable, monic and minimum-phase. The situation of correlated noise, as considered in this paper, refers to the situation that $\Phi_v(\omega)$ and $H$ are non-diagonal, while we assume that we know a priori which entries of $\Phi_v$ are nonzero.
\end{itemize}
We will assume that the standard regularity conditions on the data are satisfied that are required for convergence results of the prediction error identification method\footnote{See \cite{Ljung:99} page 249. This includes the property that $e(t)$ has bounded moments of order higher than $4$.}.

When combining the $L$ node signals we arrive at the full network expression
\begin{align*}
\begin{bmatrix}  \! w_1 \!  \\[1pt] \! w_2 \!  \\[1pt]  \! \vdots \! \\[1pt] \! w_L \!  \end{bmatrix} \!\!\! = \!\!\!
\begin{bmatrix}
0 &\! G_{12} \!& \! \cdots \! &\!\! G_{1L} \!\\
\! G_{21} \!& 0 & \! \ddots \! &\!\!  \vdots \!\\
\vdots &\! \ddots \!& \! \ddots \! &\!\! G_{L-1 \ L} \!\\
\! G_{L1} \!&\! \cdots \!& \!\! G_{L \ L-1} \!\! &\!\! 0
\end{bmatrix} \!\!\!\!
\begin{bmatrix} \! w_1 \!\\[1pt]  \! w_2 \!\\[1pt] \! \vdots \!\\[1pt] \! w_L \! \end{bmatrix} \!\!\!
+ \!\!
\begin{bmatrix} \! u_1 \!\\[1pt] \! u_2 \!\\[1pt] \! \vdots \!\\[1pt]  \! u_{L} \!\end{bmatrix}
\!\!\!+\!\!
H  \!\! \begin{bmatrix}\! e_1 \!\\[1pt] \! e_2 \!\\[1pt] \! \vdots \!\\[1pt] \! e_L\!\end{bmatrix} \!\!\!
\end{align*}
which results in the matrix equation:
\begin{align} \label{eq.dgsMatrix}
w = G w + Rr + H e.
\end{align}
It is assumed that the dynamic network is stable, i.e. $(I-G)^{-1}$ is stable, and well posed (see \cite{Dankers_diss:14} for details).
The representation (\ref{eq.dgsMatrix}) is an extension of the dynamic structure function representation \cite{Goncalves&Warnick:08}.
The identification problem to be considered is the problem of identifying one particular module $G_{ji}(q)$ on the basis of a selection of measured variables $w$, and possibly r.

Let us define $\mathcal{N}_j$ as the set of node indices $k$ such that $G_{jk} \neq 0$, i.e. the node signals in $\mathcal{N}_j$  are the \emph{$w$-in-neighbors of the node signal $w_j$}.
Let $\mathcal{D}_j$ denote the set of indices of the internal variables that
are chosen as predictor inputs. It seems most obvious to have $\mathcal{D}_j\subset\mathcal{N}_j$, but this is not necessary, as will be shown later in this paper.
Let
$\mathcal{V}_j$ denote the set of node indices $k$ such that $v_k$ has a path to $w_j$.
Let
$\mathcal{Z}_j$ denote the set of indices not in $\{j\} \cup \mathcal{D}_j$, i.e. $\mathcal{Z}_j = \{1, \ldots, L\} \setminus \{ \{j\} \cup \mathcal{D}_j \}$, reflecting the node signals that are discarded in the prediction/identification.  Let $w_{\smallsub{D}}$ denote the vector $[w_{k_1} \ \cdots \
w_{k_n} ]^T$, where $\{k_1,\ldots,k_n \} = \mathcal{D}_j$. Let
$u_{\smallsub{D}}$ denote the vector $[u_{k_1} \ \cdots \ u_{k_n}]^T$, where $\{k_1,\ldots,k_n\} = \mathcal{D}_j$. The vectors
$w_{\smallsub{Z}}$, $v_{\smallsub{D}}$, $v_{\smallsub{Z}}$ and
$u_{\smallsub{Z}}$ are defined analogously. The ordering of the
elements of $w_{\smallsub{D}}$, $v_{\smallsub{D}}$, and
$u_{\smallsub{D}}$ is not important, as long as it is the same for
all vectors. The transfer function
matrix between $w_{\smallsub{D}}$ and $w_j$ is denoted
$G_{j\smallsub{D}}$. The other transfer function
matrices are defined analogously.

\begin{table}[h]
\caption{Table with notation of variables and sets.}
\begin{tabular}{ll}
$G$             & Network matrix with modules \\
$H$             & Network noise model \\
$G_{ji}$        & Target module with input $w_i$ and output $w_j$\\
$w_j$           & Node signal $w_j$, output of the target module\\
$w_i$           & Node signal $w_i$, input of the target module \\
$\mY$           & Set of indexes of nodes that appear in the vector of predicted \\ & outputs\\
$\mD$    & Set of indexes of nodes that appear in the vector of predictor \\ & inputs for predicted outputs $w_{\smY}$\\
$\mD_j$         & Set of indexes of nodes that appear in the vector of predictor \\ & inputs for prediction of node $w_j$\\
$w_o$           & Output node signal $w_j$ if it is not in set $w_{\smQ}$\\
$\mQ$           & Set of indexes of nodes that appear both in the predicted output, \\ & and in the predictor input\\
$\mU$           & Set of indexes of nodes that only appear as predictor input: \\ & $\mU = \mD \backslash \mQ$\\
$\mA$           & Set of indexes of nodes that only appear as predictor input, \\ & that do not have any confounding variable effect: $\mA \subseteq \mU$\\
$\mB$           & Set of indexes of nodes that only appear as predictor input: \\ & $\mB = \mU\backslash\mA$\\
$\mZ$           & Set of indexes of nodes that are removed (immersed) from the \\ & network when predicting $w_{\smY}$\\
$\mZ_j$           & Set of indexes of nodes that are removed (immersed) from the \\ & network when predicting $w_j$\\
$v_k$           & Disturbance signal on node $w_k$\\
$\mathcal{N}_j$ & Index set of nodes that are $w$-in-neighbors of $w_j$\\
$e$             & (White noise) innovation of the noise process $v$\\
$\mL$           & Index set of all node signals: $[1,L]$\\
$\bar G$        & Network matrix of the immersed and transformed network (\ref{eq5})\\
$\xi$            & (White noise) innovation of the noise process in the immersed \\ & and transformed network (\ref{eq5})\\
\end{tabular}
\end{table}

\begin{figure}[htb]
\centerline{\includegraphics[scale=0.5]{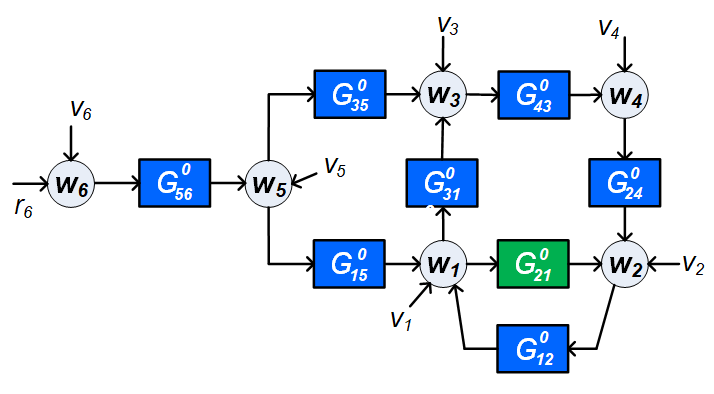}}
\vspace{-0.35cm}
	\caption{Example network with target module $G_{21}^0$ (in green).}
	\label{fig1}
\end{figure}

To illustrate the notation, consider the network sketched in Figure \ref{fig1}, and let module $G_{21}^0$ be the target module for identification.
\vspace{-0.41cm}
Then $j=2$, $i=1$; $\mathcal{N}_j = \{1,4\}$. If we choose the set of predictor inputs as $\mathcal{D}_j = \mathcal{N}_j$, then the set of remaining (nonmeasured) signals, becomes $\mathcal{Z}_j = \{ 3,5,6\}$.

By this notation, the network equation (\ref{eq.dgsMatrix}) is rewritten as:
\begin{align} \label{eq.dgsPartitioned}
\begin{bmatrix} w_j \\ w_{\smallsub{D}} \\
w_{\smallsub{Z}} \end{bmatrix} =
\begin{bmatrix} 0 & G_{j \smallsub{D}} & G_{j \smallsub{Z}} \\
G_{\smallsub{D} j} & G_{\smallsub{D} \smallsub{D}} & G_{\smallsub{D}
	\smallsub{Z}} \\ G_{\smallsub{Z} j} &
G_{\smallsub{Z} \smallsub{D}} & G_{\smallsub{Z}
	\smallsub{Z}} \end{bmatrix}
\begin{bmatrix} w_j \\ w_{\smallsub{D}} \\
w_{\smallsub{Z}} \end{bmatrix} +
\begin{bmatrix} v_j \\ v_{\smallsub{D}} \\
v_{\smallsub{Z}} \end{bmatrix} +
\begin{bmatrix} u_j \\ u_{\smallsub{D}} \\
u_{\smallsub{Z}} \end{bmatrix},
\end{align}
where $G_{\smallsub{D} \smallsub{D}}$ and $G_{\smallsub{Z}
	\smallsub{Z}}$ have zeros on the diagonal.

For identification of module $G_{ji}$ we select $\mathcal{D}_j$ such that $i \in \mathcal{D}_j$, and subsequently estimate a multiple-input single-output model for the transfer functions in $G_{j\smallsub{D}}$, by considering the one-step-ahead predictor\footnote{$\Eb$ refers to $\lim_{N\rightarrow\infty} \frac{1}{N} \sum_{t=1}^N \E$, and $w_j^{\ell}$ and $w_{\mathcal{D}_j}^{\ell}$ refer to signal samples $w_j(\tau)$ and $w_k(\tau)$, $k\in \mathcal{D}_j$, respectively, for all $\tau \leq \ell$.}
$\hat w_j(t|t-1;\theta):= \Eb \{w_j(t)\ |\ w_j^{t-1},w_{\mathcal{D}_j}^t;\theta\}$ (\cite{Ljung:99})
and the resulting prediction error $\varepsilon_j(t,\theta) = w_j(t) - \hat{w}_j(t|t-1;\theta)$, leading to:
%
\beq \label{eq.predictionError}
\varepsilon_j(t,\theta) =  H_j(\theta)^{-1} [( w_j - \!\! \sum_{k \in \mathcal{D}_j}
G_{jk}(\theta) w_k -  u_j]
\eeq
\noindent where arguments $q$ and $t$ have been dropped for notational
clarity. The parameterized transfer functions $G_{jk}(\theta)$, $k \in
\mathcal{D}_j$ and
$H_j(\theta)$ are estimated by
minimizing the sum of squared (prediction) errors:
$V_j(\theta) = \frac{1}{N}
\sum_{t=0}^{N-1} \varepsilon_j^2(t,\theta),
$
where $N$ is the length of the data set. We refer to this identification method as the {\it direct method}, \cite{VandenHof&etal_Autom:13}.

\medskip

\section{Available results and problem specification}
\label{sec:avail}

The following results are available from previous work:
\begin{itemize}
	\item When $\mathcal{D}_j$ is chosen equal to $\mathcal{N}_j$ and noise $v_j$ is uncorrelated to all $v_k$, $k\in\mathcal{V}_j$, then $G_{ji}$ can be consistently estimated in a MISO setup, provided that there is enough excitation in the predictor input signals, see \cite{VandenHof&etal_Autom:13}.
	\item When $\mathcal{D}_j$ is a subset of $\mathcal{N}_j$, and disturbance are uncorrelated, confounding variables\footnote{A confounding variable is an unmeasured variable that has paths to both the input and output of an estimation problem \cite{Pearl:2000}.} can occur in the estimation problem, and these have to be taken into account in the choice of $\mathcal{D}_j$ in order to arrive at consistent estimates of $G_{ji}$, see \cite{Dankers&etal_TAC:16}.
	\item In \cite{Dankers&etal_IFAC:17} relaxed conditions for the selection of $\mD_j$ have been formulated, while still staying in the context of MISO identification with noise spectrum of $v$ ($\Phi_v$) being diagonal. This is particularly done by choosing additional predictor input signals that are not in $\mathcal{N}_j$,.i.e. that are no in-neighbors of the output $w_j$ of the target module.
	\item For non-diagonal $\Phi_v$, an indirect/two-stage identification method can be used to arrive at consistent estimates of $G_{ji}$ \cite{VandenHof&etal_Autom:13,Dankers&etal_TAC:16,Gevers&etal:sysid18}. However the drawback of these methods is that they do not allow for a maximum likelihood analysis, i.e. they will not lead to minimum variance results.
\item This latter argument also holds for the method in \cite{Materassi&Salapaka:CDC15,Materassi&Salapaka:20}, where Wiener-filter estimates are combined to provide local module estimates, and diagonal $\Phi_v$ is considered.
\end{itemize}
In this paper, we go beyond consistency properties, and address the following problem: How to identify a single module in a dynamic network for the situation that the disturbance signals can be correlated, i.e. $\Phi_v$ not necessarily being diagonal, such that the estimate is consistent and asymptotically has Maximum Likelihood, and thus also minimum variance, properties.
%
%
\begin{figure}[htb]
	\centerline{\includegraphics[scale=0.6]{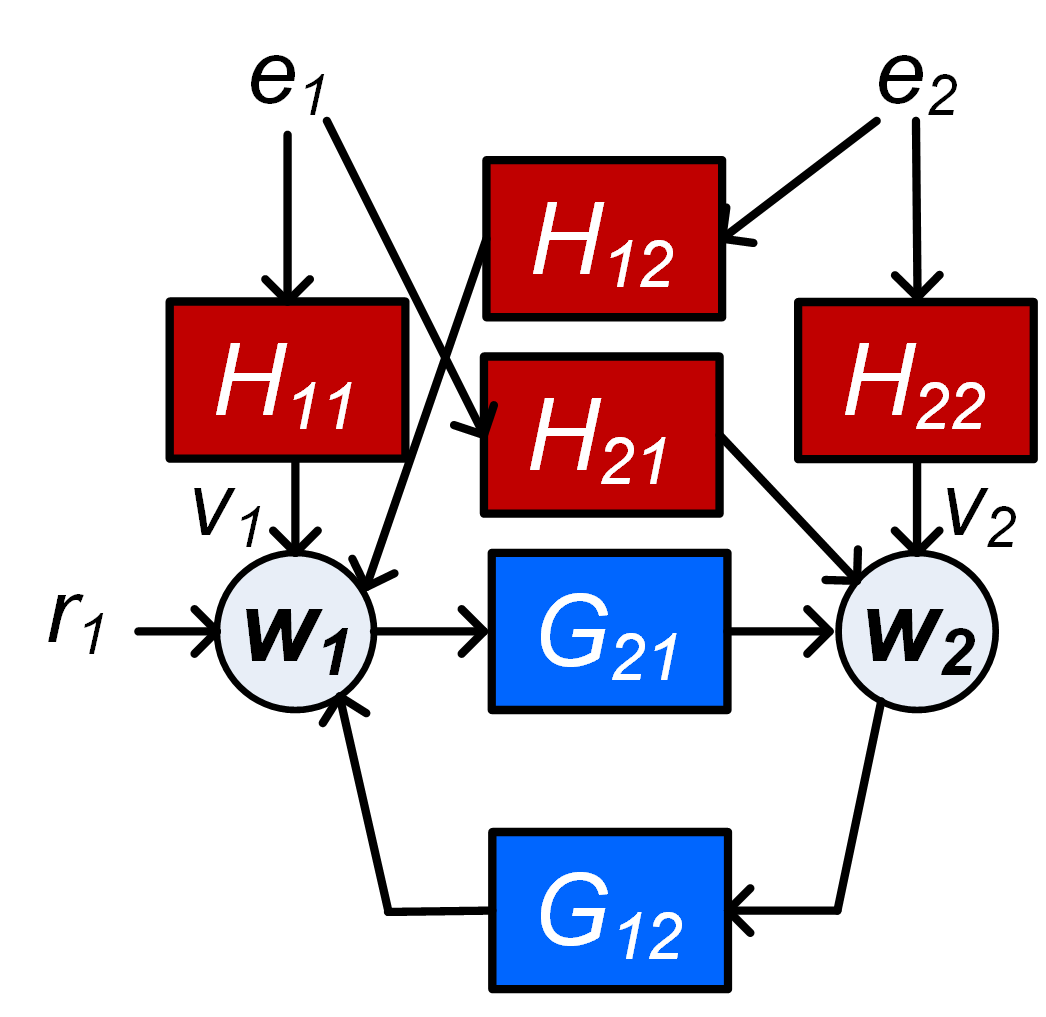}}
	\vspace{-0.35cm}
	\caption{Two-node example network from \cite{VandenHof&etal_CDC:17} with $v_1$ and $v_2$ dynamically correlated and $e_1$, $e_2$ white noise processes.}
	\label{fig:twonode}
\end{figure}
Addressing this problem requires a more careful treatment and modelling of the noise that is acting on the different node signals. This can be illustrated through a simple Example that is presented in \cite{VandenHof&etal_CDC:17}, where a two-node network is considered as given in Figure \ref{fig:twonode}, with $v_1$ and $v_2$ being dynamically correlated. In this case, a SISO identification using the direct method with input $w_1$ and output $w_2$ will lead to a biased estimate of $G_{21}$ because of the unmodelled correlation
of the disturbance signals on $w_1$ and $w_2$\footnote{In this particular example the bias is caused by the presence of $H_{21}$.}. For an analysis of this, see \cite{VandenHof&etal_CDC:17}. If both node signals $w_1$ and $w_2$ are predicted as outputs, then the correlation between the disturbance signals can be incorporated in a $2\times2$ non-diagonal noise model, thus leading to an unbiased estimate of $G_{21}$. In this way bias due to correlation in the noise signals can be avoided by predicting additional outputs other than the output of the target module. This leads to the following two suggestions:

\begin{itemize}
	\item confounding variables can be dealt with by modelling correlated disturbances on the node signals, and
	\item this can be done by moving from a MISO identification setup to a MIMO setup.
\end{itemize}
These suggestions are being explored in the current paper.
Next we will present an example to further illustrate the problem.

\begin{example}\label{exam1}
	Consider the network sketched in Figure \ref{fig1}, and let module $G_{21}^0$ be the target module for identification.
If the node signals $w_1$, $w_2$ and $w_4$ can be measured, then a two-input one-output model with inputs $w_1, w_4$ and output $w_2$ can be considered. This can lead to a consistent estimate of $G_{21}^0$ and $G_{24}^0$, provided that the disturbance signal $v_2$ is uncorrelated to all other disturbance signals.
However if e.g. $v_4$ and $v_2$ are dynamically correlated, implying that a noise model $H$ of the two-dimensional noise process is non-diagonal, then a biased estimate will result for this approach. A solution is then to include $w_4$ in the set of predicted outputs, and by adding node signal $w_3$ as predictor input for $w_4$. We then combine predicting $w_2$ on the basis of $(w_1, w_4)$ with predicting $w_4$ on the basis of $w_3$. The correlation between $v_2$ and $v_4$ is then covered by modelling a $2\times2$ non-diagonal noise model of the joint process $(v_2, v_4)$.
\end{example}

In the next sections we will formalize the procedure as sketched in Example \ref{exam1} for general networks.

\section{Concepts and notation}
\label{sec:concepts}
\noindent In line with \cite{Pearl:2000} we define the notion of confounding variable.
\begin{definition}[confounding variable] \label{def1}
	Consider a dynamic network defined by
	\beq
	\label{eqsys1}
	w = Gw + He + u
	\eeq
	with
$e$ a white noise process,
and consider the graph related to this network, with node signals $w$  and $e$. Let $w_{\smX}$ and $w_{\smY}$ be
	two subsets of measured node signals in $w$, and let $w_{\smZ}$ be the set of unmeasured node signals in $w$.
Then a noise component $e_{\ell}$ in $e$ is a {\it confounding variable for the estimation problem $w_{\smX} \rightarrow w_{\smY}$}, if in the graph there exist simultaneous paths\footnote{A simultaneous path from $e_1$ to node signal $w_1$ and $w_2$ implies that there exist a path from $e_1$ to $w_1$ as well as from $e_1$ to $w_2$.} from $e_{\ell}$ to node signals $w_{k}, k \in \mX$ and $w_{n}, n \in \mY$, while these paths are either direct\footnote{A direct path from $e_1$ to node signal $w_1$ implies that there exist a path from $e_1$ to $w_1$ which does not pass through nodes in $w$.} or only run through nodes that are in $w_{\smZ}$. \hfill $\Box$
\end{definition}

We will denote $w_{\smY}$ as the node signals in $w$ that serve as predicted outputs, and $w_{\smD}$ as the node signals in $w$ that serve as predictor inputs. Next we decompose $w_{\smY}$ and $w_{\smD}$ into disjoint sets according to: $ \mY  =  \mQ \cup \{o\} \ ; \ \mD  =  \mQ \cup \mU$ where $w_{\smQ}$ are the node signals that are common in $w_{\smY}$ and $w_{\smD}$; $w_o$ is the output $w_j$ of the target module; if $j \in \mQ$ then $\{o\}$ is void; $w_{\smU}$ are the node signals that are only in $w_{\smD}$. In this situation the measured nodes will be $w_{\smD\cup\smY}$ and the unmeasured nodes $w_{\smZ}$ will be determined by the set
$\mZ = \mL \backslash \{ \mD \cup \mY\}$, where $\mL = \{1,2,\cdots L\}$.
\begin{figure}[h]
	\centerline{\includegraphics[scale=0.5]{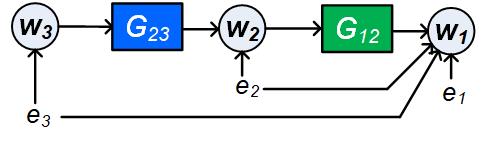}}
	\caption{A simple network with 3 nodes $w_1$, $w_2$, $w_3$ and unmeasured noise sources $e_1$, $e_2$ and $e_3$. $G_{12}$ is the target module to be identified.}
	\label{fig11}
\end{figure}
There can exist two types of confounding variable namely \emph{direct and indirect confounding variables}. For \emph{direct confounding variables} the simultaneous paths mentioned in the definition are both \emph{direct paths}, while in all other cases we refer to the confounding variables as \emph{indirect confounding variables}.
For example, in the network as shown in Figure \ref{fig11} with $\mD = \{2\}$, $\mY = \{1\}$ and $\mZ = \{3\}$, for the estimation problem $w_2 \rightarrow w_1$, $e_2$ is a \emph{direct confounding variable} since it has a simultaneous path to $w_1$ and $w_2$ where both the paths are \emph{direct paths}. Meanwhile $e_3$ is an \emph{indirect confounding variable} since it has a simultaneous path to $w_1$ and $w_2$ where one of the path is an unmeasured path\footnote{An unmeasured path is a path that runs through nodes in $w_{\smZ}$ only. Analogously, we can define unmeasured loops through a node $w_k$.}.
%
\begin{remark}
Confounding variables are defined in accordance with their use in \cite{Dankers&etal_IFAC:17}, on the basis of a network description as in (\ref{eqsys1}). In this definition absence of confounding variables still allows that there are unmeasured signals that create correlation between the inputs and outputs of an estimation problem, in particular if the white noise signals in $e$ are statically correlated, i.e $cov(e)$ being non-diagonal. It will appear that this type of correlations will not hinder our identification results, as analysed in Section \ref{sec:ident}.
\end{remark}
%
\section{Main results - Line of reasoning}
\label{sec:Main}
%
On the basis of the decomposition of node signals as defined in the previous section we are going to represent the system's equations (\ref{eqsys1}) in the following structured form:
\beqr
\label{eq1}
\begin{bmatrix} w_{\smQ} \\ w_o \\ w_{\smU} \\ w_{\smZ} \end{bmatrix}\! &\! =\! &\!
\begin{bmatrix} G_{\smQ\smQ} & G_{\smQ o} & G_{\smQ\smU} & G_{\smQ\smZ} \\
	G_{o\smQ} & G_{oo} & G_{o\smU} & G_{o\smZ} \\
	G_{\smU\smQ} & G_{\smU o} & G_{\smU\smU} & G_{\smU\smZ} \\
	G_{\smZ\smQ} & G_{\smZ o} & G_{\smZ\smU} & G_{\smZ\smZ} \end{bmatrix}\!
\begin{bmatrix} w_{\smQ} \\ w_o \\ w_{\smU} \\ w_{\smZ} \end{bmatrix}\! +\! R(q)r  \nonumber \\
& & + \begin{bmatrix} H_{\smQ\smQ} & H_{\smQ o} & H_{\smQ\smU} & H_{\smQ\smZ} \\
	H_{o\smQ} & H_{oo} & H_{o\smU} & H_{o\smZ} \\
	H_{\smU\smQ} & H_{\smU o} & H_{\smU\smU} & H_{\smU\smZ} \\
	H_{\smZ\smQ} & H_{\smZ o} & H_{\smZ\smU} & H_{\smZ\smZ} \end{bmatrix}\!
\begin{bmatrix} e_{\smQ} \\ e_o \\ e_{\smU} \\ e_{\smZ} \end{bmatrix}
\eeqr
where we make the notation agreement that the matrix $H$ is not necessarily monic, and the scaling of the white noise process $e$ is such that $cov(e) = I$. Without loss of generality, we can assume $r = 0$ for the sake of brevity.

\begin{figure}[h]
	\centerline{\includegraphics[scale=0.42]{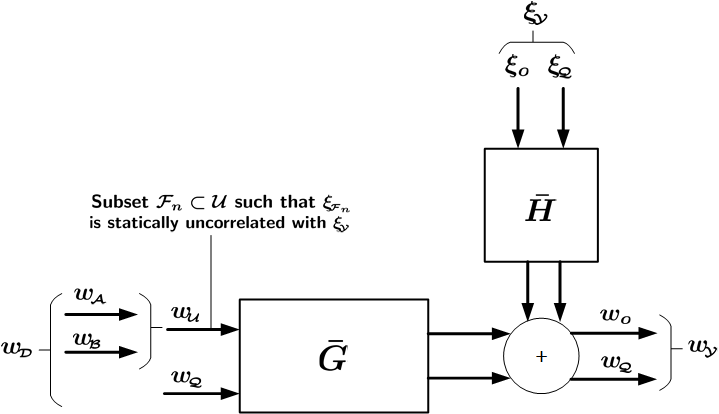}}
	\caption{Figure to depict the identification setup and classification of different sets of signals in the input and output of the identification problem.}
	\label{fig11a}
\end{figure}

Our objective is to end up with an an identification problem in which we identify the dynamics from inputs $(w_{\smQ},w_{\smU})$ to outputs $(w_{\smQ},w_o)$, while our target module $G_{ji}(q)$ is present as one of the scalar transfers (modules) in this identified (MIMO) model. This can be realized by the following steps:

\begin{enumerate}
\item Firstly, we write the system's equations for the measured variables as
\beq
\label{eq5a}
\underbrace{\begin{bmatrix} w_{\smQ} \\ w_o \\ \hline w_{\smU} \end{bmatrix}}_{w_m}\! =\!
\underbrace{\left[ \begin{array}{c|c} \bar G & 0 \\ \hline \\[-3mm] \bar G_{\smU\smD} & \bar G_{\smU o} \end{array} \right]}_{\bar G_m}\!
\begin{bmatrix} w_{\smQ} \\ w_{\smU} \\ \hline w_o \end{bmatrix} +
\underbrace{\left[ \begin{array}{c|c} \bar H & 0 \\ \hline \\[-3mm] 0 & \bar H_{\smU\smU} \end{array} \right]}_{\bar H_m}\!
\underbrace{\begin{bmatrix} \xi_{\smQ} \\ \xi_o \\ \hline \xi_{\smU} \end{bmatrix}}_{\xi_m}
\eeq
with $\xi_m$ a white noise process, while $\bar H$ is monic, stable and stably invertible and the components in $\bar G$ are zero if it concerns a mapping between identical signals. This step is made by removing the non-measured signals $w_{\smZ}$ from the network, while maintaining the second order properties of the remaining signals. This step is referred to as immersion of the nodes in $w_{\smZ}$ \cite{Dankers&etal_TAC:16}.
\item As an immediate result of the previous step we can write an expression for the output variables $w_{\smY}$, by considering the upper part of the equation (\ref{eq5a}), as
\beq
\label{eq5}
\underbrace{\begin{bmatrix} \!w_{\smQ}\! \\ \!w_o\! \end{bmatrix}}_{w_{\smY}}\!\! =\!\!
\underbrace{\begin{bmatrix} \!\bar G_{\smQ\smQ}\! &\! \bar G_{\smQ\smU}\!  \\
		\!\bar G_{o\smQ}\! & \!\bar G_{o\smU}\! \end{bmatrix}}_{\bar G}
\underbrace{\begin{bmatrix} \!w_{\smQ}\! \\ \!w_{\smU}\! \end{bmatrix}}_{w_{\smD}} +
\underbrace{\begin{bmatrix} \bar H_{\smQ\smQ} & \bar H_{\smQ o} \\
		\bar H_{o\smQ} & \bar H_{oo} \end{bmatrix}}_{\bar H}
\underbrace{\begin{bmatrix} \xi_{\smQ} \\ \xi_o \end{bmatrix}}_{\xi_{\smY}}
\eeq
with $cov(\xi_{\smY}) := \bar\Lambda$.
\item Thirdly, we will provide conditions to guarantee that $\bar G_{ji}(q) = G_{ji}(q)$, i.e the target module appearing in equation (\ref{eq5}) is the target module of the original network \emph{(invariance of target module)}. This will require conditions on the selection of node signals in $w_{\smQ}, w_{o},w_{\smU}$.
\item Finally, it will be shown that, on the basis of (\ref{eq5}), under fairly general conditions, the transfer functions $\bar G(q)$ and $\bar H(q)$ can be estimated consistently, and with maximum likelihood properties. A pictorial representation of the identification setup with the classification of different sets of signals in \eqref{eq5} is provided in Figure \ref{fig11a}. The figure also contains set $\mA, \mB, \mF_n$ which will be introduced in the sequel.
\end{enumerate}

\vspace*{-4mm}
The combination of steps 3 and 4 will lead to a consistent and maximum likelihood estimation of the target module $G_{ji}(q)$.
It has to be noted that an identification setup results, in which signals can simultaneously act as input and as output (the set $w_{\smQ}$). Because $\bar G_{\smQ\smQ}$ is restricted to be hollow, this does not lead to trivial transfers between signals that are the same. A related situation appears when identifying  a full network, while using all node signals as both inputs and outputs, as in \cite{Weerts&etal_Autom:18_reducedrank}.

The steps 1)-4) above will require conditions on the selection of node signals, based on the known topology of the network and an allowed correlation structure of the disturbances in the network. Specifying these conditions on the selection of sets $w_{\smQ}, w_{o}, w_{\smU}$, will be an important objective of the next section.

\section{Main Results - Derivations}

\subsection{System representation after immersion (Steps 1-2)}

First we will show that a network in which signals in $w_{\smZ}$ are removed (immersed) can indeed be represented by
(\ref{eq5a}).

\begin{proposition}
\label{propx1}
Consider a dynamic network given by (\ref{eq1}),
where the set of all nodes $w_{\smL}$ is decomposed in disjunct sets $w_{\smQ}$, $w_o$, $w_{\smU}$ and $w_{\smZ}$ as defined in Section \ref{sec:concepts}. Then, for the situation $r=0$,
\begin{enumerate}
\item
there exists a representation (\ref{eq5a}) of the measured node signals $w_m$, with $\bar H_m$ monic, stable and stably invertible, and $\xi_m$ a white noise process, and
\item for this representation there are no confounding variables for the estimation problem $w_{\smU} \rightarrow w_{\smY}$.
\end{enumerate}

\end{proposition}

{\bf Proof}: See appendix.

\begin{figure*}[hbt]
\hspace*{3cm} (a) \hspace*{5cm} (b) \hspace*{55mm} (c) \\

	\centerline{\includegraphics[scale=0.36]{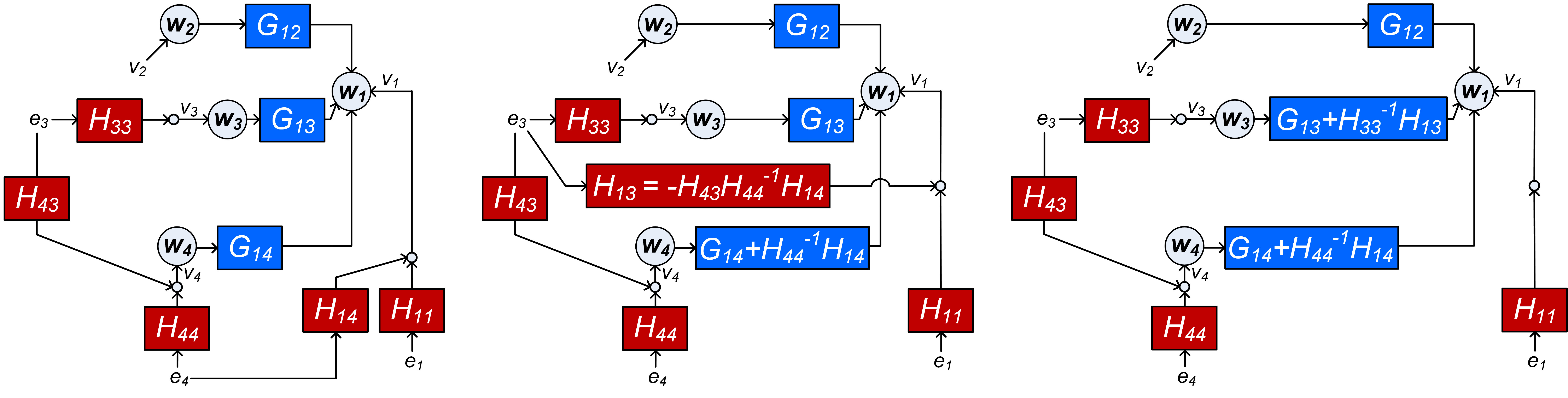}}
	\caption{(a): Original network with 4 nodes $\{w_i\}_{i=1,\cdots 4}$, and unmeasured white noise sources $\{e_i\}_{i=1,\cdots 4}$; (b): Transformed network with confounding variable for $w_4 \rightarrow w_1$ removed; (c): Transformed network with also the confounding variable for $w_3 \rightarrow w_1$ removed.}
	\label{fig21}
\end{figure*}

\medskip
The consequence of Proposition \ref{propx1} is that the output node signals in $w_{\smY}$ can be explicitly written in the form of (\ref{eq5}), in terms of input node signals $w_{\smD}$ and disturbances, without relying on (unmeasured) node signals in $w_{\smZ}$. The particular structure of network representation (\ref{eq5a}) implies that there are no confounding variables for the estimation problem $w_{\smU} \rightarrow w_{\smY}$. This will be an important phenomenon for our identification setup.
Based on (\ref{eq5}), a typical prediction error identification method can provide estimates of $\bar G$ and $\bar H$ from measured signals $w_{\smY}$ and $w_{\smD}$ with $\mD = \mQ\cup\mU$. In this estimation problem, confounding variables for the estimation problem $w_{\smQ} \rightarrow w_{\smY}$ are treated by correlated noise modelling in $\bar H$, while confounding variables for the estimation problem $w_{\smU} \rightarrow w_{\smY}$ are not present, due to the structure of (\ref{eq5a}).

In the following example, the step towards (\ref{eq5a}) will be illustrated, as well as its effect on the dynamics in $\bar G$.

\begin{example}
\label{exprop1}
Consider the 4-node network depicted in Figure \ref{fig21}(a), where all nodes are considered to be measured, and where we select $w_o = w_1$, $\mU = \{2,3,4\}$, and $\mQ = \emptyset$. In this network, there is a confounding variable $e_4$ for the problem $w_{4} \rightarrow w_{1}$ (i.e $w_{\smU} \rightarrow w_{\smY}$), meaning that for the situation $\xi=e$ the noise model $\bar H_m$ in (\ref{eq5a}) will not be block diagonal. Therefore the network does not comply with the representation in \eqref{eq5a} and \eqref{eq5}. We can remove the confounding variable, by shifting the effect of $H_{14}$ into a transformed version of $G_{14}$, which now becomes $G_{14}+H_{44}^{-1}H_{14}$, as depicted in Figure \ref{fig21}(b). However, since this shift also affects the transfer from $e_3$ to $w_1$, the change of $G_{14}$ needs to be mitigated by a new term $H_{13}$, in order to keep the network signals invariant. In the resulting network the confounding variable for $w_4 \rightarrow w_1$ is removed, but a new confounding variable ($e_3)$ for $w_3 \rightarrow w_1$ has been created. In the second step, shown in Figure \ref{fig21}(c), the term $H_{13}$ is removed by incorporating its effect in the module $G_{13}$ which now becomes $G_{13}+H_{33}^{-1}H_{13}$. In the resulting network there are no confounding variables for $w_{\smU} \rightarrow w_1$. This representation complies with the structure in (\ref{eq5a}). Note that in the transformed network, the dynamics of $G_{12}$ is left invariant, while the dynamics of $G_{14}$ and $G_{13}$ have been changed. The intermediately occurring confounding variables relate to a {\it sequence of linked confounders}, as discussed in \cite{Dankers&etal_IFAC:17}.
\hfill $\Box$
\end{example}

In the next subsection it will be investigated under which conditions our target module will remain invariant under the above transformation to a representation (\ref{eq5a}) without confounding variables.

\subsection{Module invariance result (Step 3)}
\label{sec:modinv}

The transformation of a network into the form (\ref{eq5a}), leading to the resulting identification setup of (\ref{eq5}), involves two basic steps, each of which can lead to a change of dynamic modules in $\bar G$. These two steps are
\begin{enumerate}
\item[(a)] Removing of non-measured signals in $w_{\smZ}$ (immersion), and
\item[(b)] Transforming the system's equations to a form where there are no confounding variables for $w_{\smU} \rightarrow w_{\smY}$.
\end{enumerate}
Module invariance in step (a) is covered by the following Condition:
\begin{condition}[parallel path and loop condition\cite{Dankers&etal_TAC:16}]
\label{condx1}
Let $G_{ji}$ be the target network module to be identified.
In the original network (\ref{eq1}):
\begin{itemize}
\item Every path from $w_i$ to $w_j$, excluding the path through $G_{ji}$, passes through a node $w_k, k \in \mD$, and
\item Every loop through $w_j$ passes through a node in $w_k, k \in \mD$. \hfill $\Box$
\end{itemize}
\end{condition}
This condition has been introduced in \cite{Dankers&etal_TAC:16} for a MISO identification setup, to guarantee that when immersing (removing) nonmeasured node signals from the network, the target module will remain invariant. As an alternative, more generalized notions of network abstractions have been developed for this purpose in \cite{Weerts&etal_Autom:19_abstraction}. Condition \ref{condx1} will be used to guarantee module invariance under step (a).

\medskip
Step (b) above is a new step, and requires studying module invariance in the step transforming a network from an original format where all nodes are measured, into a structure that complies with (\ref{eq5a}), i.e. with absence of confounding variables for
$w_{\smU} \rightarrow w_{\smY}$.

We are going to tackle this problem, by decomposing the set $\mU$ into two disjunct sets
$\mU = \mA \cup \mB$
aiming at the situation that in the transformed network, the modules $G_{\smY\smA}$ stay invariant, while for the modules $G_{\smY\smB}$ we accept that the transformation can lead to module changes.
We construct $\mA$ by choosing signals $w_k \in w_{\smU}$ such that in the original network there are no confounding variables for the estimation problem $w_{\smA} \rightarrow w_{\smY}$. For the selection of $\mB$, we do allow confounding variables for the estimation problem $w_{\smB} \rightarrow w_{\smY}$. By requiring a particular ``disconnection'' between the sets $\mA$ and $\mB$, we can then still guarantee that the modules $G_{\smY\smA}$ stay invariant.


The following condition will address the major requirement for addressing our step (b).

\begin{condition}
\label{condx2a}
$\mU$ is decomposed into two disjunct sets, $\mU = \mA \cup \mB$ (see Figure \ref{fig11a}), such that \emph{in the original network} (\ref{eq1}) there are no confounding variables for the estimation problems $w_{\smA} \rightarrow w_{\smY}$ and $w_{\smA} \rightarrow w_{\smB}$. \hfill $\Box$
\end{condition}

Condition \ref{condx2a} is not a restriction on $\mU$, as such a decomposition can always be made, e.g. by taking $\mA = \emptyset$ and $\mB = \mU$. The flexibility in choosing this decomposition will be instrumental in the sequel of this paper.

\begin{example}[Example \ref{exprop1} continued]
\label{expropab}
In the example network depicted in Figure \ref{fig21}, we observe that in the original network there is a confounding variable for $w_{4} \rightarrow w_1$. However in the step towards creating a network without confounding variables for $w_{\smU} \rightarrow w_{\smY}$ an intermediate step occurs, where there is also a confounding variable for $w_3 \rightarrow w_1$, as depicted in Figure \ref{fig21}(b). For $\mU = \{2,3,4\}$ the choice $\mA = \{2,3\}$, $\mB = \{4\}$, is not valid
since there exists a confounding variable ($e_3$) for $w_3 \rightarrow w_4$ which violates the second condition that there should be no confounding variables for $w_{\smA} \rightarrow w_{\smB}$. Therefore the appropriate choice satisfying Condition \ref{condx2a} is $\mA = \{2\}$ and $\mB = \{3,4\}$.
Note that this matches with the situation that in the transformed network (Figure \ref{fig21}(c)), the module $G_{\smY\smA}$ remains invariant, and the modules $G_{\smY\smB}$ get changed. \hfill $\Box$
\end{example}

We can now formulate the module invariance result.

\begin{theorem}[Module invariance result]
\label{theomodinv}
Let $G_{ji}$ be the target network module. In the transformed system's equation (\ref{eq5}), it holds that $\bar G_{ji} = G^0_{ji}$ under the following conditions:
\begin{enumerate}
\item The parallel path and loop Condition \ref{condx1} is satisfied, and
\item The following three conditions are satisfied:
\begin{itemize}
\item[a.] $\mU$ is decomposed in $\mA$ and $\mB$, satisfying Condition \ref{condx2a}, and
\item[b.] $i \in \{\mA \cup \mQ\}$, and
\item[c.] Every path from $\{w_i,w_j\}$ to $w_{\smB}$ passes through a measured node in $w_{\mathcal{L}\backslash\mZ}$.
\end{itemize}
\end{enumerate}
\end{theorem}
{\bf Proof}: See appendix.

A more detailed illustration of the conditions in the theorem will be deferred to three different algorithms for selecting the node signals, to be presented in Sections \ref{sec:fullinput}-\ref{sec:usersel}. We will first develop the identification results for the general case.

\subsection{Identification results (Step 4)}
\label{sec:ident}
If the conditions of Theorem \ref{theomodinv} are satisfied, then the target module $\bar G_{ji}=G_{ji}^0$ can be identified on the basis of the system's equation (\ref{eq5}). For this system's equation we can set up a predictor model with input $w_{\smD}$ and outputs $w_{\smY}$, for the estimation of $\bar G$ and $\bar H$. This will be based on a parameterized model set determined by
\[ \M := \left\{(\bar G(\theta), \bar H(\theta), \bar\Lambda(\theta)), \theta\in\Theta\right\}, \]
while the actual data generating system is represented by $\mathcal{S} = (\bar G(\theta_o), \bar H(\theta_o), \bar\Lambda(\theta_0))$.
The corresponding identification problem is defined by considering the one-step-ahead prediction of $w_{\smY}$ in the parametrized model, according to
$\hat w_{\smY}(t|t-1;\theta) := \E \{w_{\smY}(t)\ |\ w_{\smY}^{t-1}, w_{\smD}^{t};\theta \}$
where $w_{\smD}^t$ denotes the past of $w_{\smD}$, i.e. $\{w_{\smD}(k), k\leq t\}$. The resulting prediction error becomes:
$\varepsilon(t,\theta) := w_{\smY}(t) - \hat w_{\smY}(t|t-1;\theta)$, leading to
\begin{equation} \label{eqx5}
\varepsilon(t,\theta)= \bar H(q,\theta)^{-1}
\left[ w_{\smY}(t) - \bar G(q,\theta)w_{\smD}(t)\right],
\end{equation}
and the weighted least squares identification criterion
\begin{equation}\label{eqx6}
\hat\theta_N = \arg\min_{\theta} \frac{1}{N} \sum_{t=0}^{N-1} \varepsilon^T(t,\theta) W  \varepsilon(t,\theta),
\end{equation}
with $W$ any positive definite weighting matrix. This parameter estimate then leads to an estimated subnetwork $\bar G_{\smY\smD}(q,\hat\theta_N)$ and noise model $\bar H(q,\tn)$, for which consistency and minimum variance results will be formulated next.

\begin{theorem}[Consistency] \label{theorem1}
Consider a dynamic network represented by (\ref{eq5a}), and a related (MIMO) network identification setup with predictor inputs $w_{\smD}$ and predicted outputs $w_{\smY}$, according to (\ref{eq5}). Let $\mF_n \subseteq \mU$ be the set of node signals $k$ for which $\xi_k$ is statically uncorrelated with $\xi_{\smY}$\footnote{This implies that $\E [\xi_k(t)\xi_{\smY}(t)] = 0$.} and let $\mF := \mU \backslash \mF_n$. Then a direct prediction error identification method according to (\ref{eqx5})-(\ref{eqx6}), applied to a parametrized model set $\mathcal{M}$ will provide consistent estimates of $\bar G$ and $\bar H$ if:
\begin{enumerate}
		\item[a.] $\mathcal{M}$ is chosen to satisfy $\mathcal{S} \in \mathcal{M}$;
		\item[b.] $\Phi_{\kappa}(\omega) > 0 $ for a sufficiently high number of frequencies, where
		$ \kappa(t) := \begin{bmatrix} w_{\smD}^\top(t) & \xi_{\smQ}^\top(t) & w_o(t) \end{bmatrix}^\top$;\\ (data-informativity condition).
		\item[c.] The following paths/loops should have at least a delay:
		\begin{itemize}
		    \item All paths/loops from $w_{\smY\cup\smF}$ to $w_{\smY}$ in the network (\ref{eq5}) and in its parametrized model; and
		    \item For every $w_k \in \mF_n$, all paths from $w_{\smY\cup\smF}$ to $w_k$ in the network (\ref{eq5}), or all paths from $w_{k}$ to $w_{\smY}$ in the parametrized model.
		 \end{itemize} \emph{(delay in path/loop condition.)}

	\end{enumerate}
\end{theorem}

{\bf Proof:} See appendix.

The consistency theorem has a structure that corresponds to the classical result of the direct prediction error identification method applied to a closed-loop experimental setup, \cite{Ljung:99}. A system in the model set condition (a), an informativity condition on the measured data (b), and a loop delay condition (c). Note however that conditions (b) and (c) are generalized versions of the typical closed-loop case \cite{Ljung:99,VandenHof&etal_Autom:13}, and are dedicated for the considered network setup.

It is important to note that Theorem \ref{theorem1} is formulated in terms of conditions on the network in \eqref{eq5a}, which we refer to as the \emph{transformed network}. However, it is quintessential to formulate the conditions in terms of properties of signals in the \emph{original network}, represented by \eqref{eq1}.
\begin{proposition}\label{propx2}
If in the original network, $\mU$ is decomposed in two disjunct sets $\mA$ and $\mB$ satisfying Condition \ref{condx2a}, then
Condition c of Theorem \ref{theorem1} can be reformulated as:
\begin{itemize}
\item[c.] The following paths/loops should have at least a delay:
\begin{itemize}
		    \item All paths/loops from $w_{\smY\cup\smB}$ to $w_{\smY}$ in the original network (\ref{eq1}) and in the parametrized model; and
		    \item For every $w_k \in \mA$, all paths from $w_{\smY\cup\smB}$ to $w_k$ in the network (\ref{eq1}), or all paths from $w_{k}$ to $w_{\smY}$ in the parametrized model.
\end{itemize}
\end{itemize}
\end{proposition}

{\bf Proof:} See appendix.

%
%
%

\medskip
Condition (b) of Theorem \ref{theorem1} requires that there should be enough excitation present in the node signals, which actually reflects a type of identifiability property \cite{Weerts&etal_Autom:18_identifiability}. Note that this excitation condition may require that there are external excitation signals present at some locations, see also \cite{VandenHof&etal_Autom:13,Gevers&Bazanella_CDC:15,vanWaarde&Tesi&Camlibel_NECSYS:18,Weerts&etal_CDC:18,Hendrickx&Gevers&Bazanella_TAC:19,Cheng&Shi&VandenHof_CDC:19}, and \cite{VandenHof&Ramaswamy_CDC:20}, where it is shown that $dim(r) \geq |\mQ|$, with $|\mQ|$ the cardinality of $\mQ$. Since we are using a direct method for identification, excitation signals $r$ are not directly used in the predictor model, although they serve the purpose of providing excitation in the network. A first result of a generalized method where, besides node signals $w$, also signals $r$ are included in the predictor inputs, is presented in \cite{Ramaswamy&etal_CDC:19}.

Since in the result of Theorem \ref{theorem1} we arrive at white innovation signals, the result can be extended to formulate Maximum Likelihood properties of the estimate.

\begin{theorem}
	\label{theorem2}
	Consider the situation of Theorem \ref{theorem1}, and let the conditions for consistency be satisfied. Let $\xi_{\smY}$ be normally distributed, and let $\bar\Lambda(\theta)$ be parametrized independently from $\bar G(\theta)$ and $\bar H(\theta)$. Then, under zero initial conditions, the Maximum Likelihood estimate of $\theta^0$ is 	
	\beqr \label{eq:ML2}
	\hat \theta_N^{ML} & = & \arg \min_{\theta} \det \left ( \frac{1}{N} \sum_{t=1}^N \varepsilon(t,\theta)  \varepsilon^T(t,\theta) \right )
	\\
	\Lambda(\hat\theta_N^{ML}) & = & \frac{1}{N} \sum_{t=1}^N  \varepsilon(t,\hat\theta_N^{ML}) \varepsilon^{T}(t,\hat\theta_N^{ML}).
	\eeqr
\end{theorem}

{\bf Proof:} Can be shown by following a similar reasoning as in Theorem 1 of \cite{Weerts&etal_Autom:18_reducedrank}. \hfill $\Box$

\medskip
So far, we have analysed the situation for given sets of node signals $w_{\smQ}$, $w_{o}$, $w_{\smA}$, $w_{\smB}$ and $w_{\smZ}$. The presented results are very general and allow for different algorithms to select the appropriate signals and specify the particular signal sets, that will guarantee target module invariance and consistent and minimum variance module estimates with the presented local direct method.
In the next sections we will focus on formulating guidelines for the selection of these sets, such that the target module invariance property holds, as formulated in Theorem \ref{theomodinv}.
For formulating these conditions, we will consider three different situations with respect to the availability of measured node signals.
\begin{enumerate}
\item[(a)] In the \emph{Full input case}, we will assume that all in-neighbors of the predicted output signals are measured and used as predictor input;
\item[(b)] In the \emph{Minimum input case}, we will include the smallest possible number of node signals to be measured for arriving at our objective;
\item[(c)] In the \emph{User selection case}, we will formulate our results for a prior given set of measured node signals;
\end{enumerate}

\section{Algorithm for signal selection: full input case}
\label{sec:fullinput}

The first algorithm to be presented is based on the strategy that for any node signal that is selected as output, we have access to all of its $w$-in-neighbors, that are to be included as predictor inputs. This strategy will lead to an identification setup with a maximum use of measured node signals that contain information that is relevant for modeling our target module $G_{ji}$. The following strategy will be followed:
\begin{itemize}
\item We start by selecting $i\in \mD$ and $j\in \mY$;
\item Then we extend $\mD$ in such a way that all $w$-in-neighbors of $w_{\smY}$ are included in $w_{\smD}$.
\item All node signals in $w_{\smD}$ that have noise terms $v_k$, $k\in\mD$ that are correlated with any $v_{\ell}$, $\ell \in \mY$ (\emph{direct} confounding variables for $w_{\smD} \rightarrow w_{\smY}$), are included in $\mY$ too. They become elements of $\mQ$.
\item With $\mA := \mD \backslash \mQ$ it follows that by construction there are no {\it direct} confounding variables for the estimation problem $w_{\smA} \rightarrow w_{\smY}$.
\item Then we choose $w_{\smB}$ as a subset of nodes that are not in $w_{\smY}$ nor in $w_{\smA}$. This set needs to be introduced to block the \emph{indirect} confounding variables for the estimation problem $w_{\smA} \rightarrow w_{\smY}$, and will be chosen to satisfy Condition 2a and 2c of Theorem \ref{theomodinv}.
\item Every node signal $w_k$, $k\in \mA$ for which there are only indirect confounding variables and cannot be blocked by a node in $w_{\smB}$, is
\begin{itemize}
	\item moved to $\mB$ if Conditions 2a and 2c of Theorem \ref{theomodinv} are satisfied and $k\neq i$; (else)
	\item included in $\mY$ and moved to $\mQ$;
\end{itemize}

\item Finally, we define the identification setup as the estimation problem $w_{\mathcal D} \rightarrow w_{\mathcal Y}$, with $\mD = \mQ \cup \mA \cup \mB$ and $\mY = \mQ \cup \{o\}$.
\end{itemize}

Note that because all $w$-in-neighbors of $w_{\smY}$ are included in $w_{\smD}$, we automatically satisfy the parallel path and loop condition \ref{condx1}.
In order for the selection of node signals $w_{\smB}$ to satisfy the conditions of Theorem \ref{theomodinv}, we will specify the following Property \ref{p1}.
\begin{property}
	\label{p1}
	Let the node signals $w_{\smB}$ be chosen to satisfy the following properties:
	\begin{enumerate}
		\item If, in the original network, there are no confounding variables for the estimation problem $w_\mA \rightarrow w_{\smY}$, then $\mB$ is void implying that $w_{\smB}$ is not present;
		\item If, in the original network, there are confounding variables for the estimation problem $w_\mA \rightarrow w_{\smY}$, then all of the following conditions need to be satisfied:%
		\begin{itemize}
			\item[a.] For any confounding variable for the estimation problem $w_\mA \rightarrow w_{\smY}$, the unmeasured paths from the confounding variable to node signals $w_{\smA}$ pass through a node in $w_{\smB}$.
			\item[b.] There are no confounding variables for the estimation problem $w_{\smA} \rightarrow w_{\smB}$.
			\item[c.] Every path from $\{w_i,w_j\}$ to $w_{\smB}$ passes through a measured node in $w_{\mathcal{L}\backslash\mZ}$.
			\hfill $\Box$
		\end{itemize}
	\end{enumerate}
\end{property}
Property 2a) ensures that, after including $w_{\smB}$ in the set of measured signals, there are no \emph{indirect} confounding variables for the estimation problem $w_{\smA} \rightarrow w_{\smY}$, and Property 2b) guarantees that there are no confounding variables for the estimation problem $w_{\smA} \rightarrow w_{\smB}$. Together we satisfy Condition 2a) of Theorem \ref{theomodinv}. Also, Property 2c) guarantees condition 2c) of Theorem \ref{theomodinv} to be satisfied. Finally, as per the algorithm, $w_i$ can be either in $w_{\smA}$ or $w_{\smQ}$. Therefore at the end of the algorithm, we will obtain sets of signals that satisfy the conditions in Theorem \ref{theomodinv} for target module invariance.

\begin{figure}[htb]
	\includegraphics[scale=0.5]{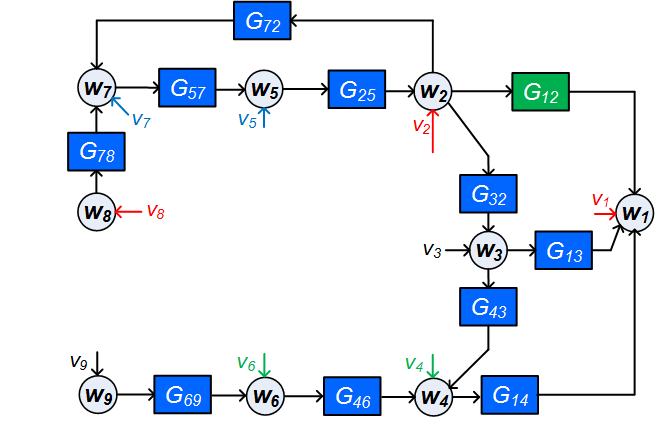}
	\caption{Example network with $v_1$ dynamically correlated with $v_2$ and $v_8$ (red colored). $v_4$ is dynamically correlated with $v_6$ (green colored) and $v_5$ is dynamically correlated with $v_7$ (blue colored).}
	\label{fig3}
\end{figure}

\begin{example}\label{exam4}
	Consider the network in Figure \ref{fig3}. $G_{12}$ is the target module that we want to identify. We now select the signals according to the algorithm presented in this section. First we include the input of the target module $w_2$ in $w_{\smD}$ and the output of the target module $w_1$ in $w_{\smY}$. Next we include all $w$-in-neighbors of $w_{\smY}$ (i.e. $w_3$ and $w_4$) in $w_{\smD}$.
All node signals in $w_{\smD}$ that have noise terms $v_k$, $k\in\mD$ that are correlated with any $v_{\ell}, \ell \in \mY$ need to be included in $\mY$ too. This concerns $w_2$, since $v_1$ is correlated with $v_2 $. Now $w_{\smY} = \{w_1,w_2\}$ has changed and we need to include the $w$-in-neighbors of $w_2$, which is $w_5$, in $w_{\smD}$, leading to $w_{\smD} = \{w_2,w_3,w_4,w_5\}$.
After a check we can conclude that all node signals in $w_{\smD}$ that have noise terms $v_k$, $k\in\mD$ that are correlated with any $v_{\ell}, \ell \in \mY$ are included in $\mY$ too. The result now becomes
	\beqr
	\mY = \{1,2\} \ &;& \ \mD = \{2,3,4,5\} \ \\
	\mQ = \mY \cap \mD = \{2\} \ &;& \ \mA = \mD \backslash \mQ = \{3,4,5\}.
	\eeqr
	Since $v_8$ is dynamically correlated with $v_1$, in the resulting situation we will have a confounding variable for the estimation problem $w_5 \rightarrow w_{1}$ (i.e. $w_{\smA} \rightarrow w_{\smY}$). As per condition 2a of Property \ref{p1}, the path of the confounding variable $e_8$ to $w_5$ should be blocked by
	a node signal in $w_{\smB}$, which can be either $w_7$ or $w_8$. $w_7$ cannot be chosen in $w_{\smB}$ since this would create a confounding variable for $w_{\smA} \rightarrow w_{\smB}$ (i.e. $w_5 \rightarrow w_7$). Moreover, $w_7 \in w_{\smB}$ would also create an unmeasured path $w_i \rightarrow w_7$ with $w_i = w_2$, thereby violating Condition 2c of Property \ref{p1}. When $w_8$ is chosen in $w_{\smB}$, the conditions in
	Property \ref{p1} are satisfied and hence we choose $\mB = \{8\}$.
The resulting estimation problem is $(w_2,w_3,w_4,w_5,w_8) \rightarrow (w_1,w_2)$, and will according to Theorem \ref{theorem1} provide a consistent and maximum likehood estimate of $G_{12}$.
\end{example}

\section{Algorithm for signal selection: minimum input case}
\label{sec:mininput}
Rather than measuring all node signals that are $w$-in-neighbors of the output $w_j$ of our target module $G_{ji}$, we now focus on an identification setup that uses a minimum number of measured node signals, according to the following strategy:
\begin{itemize}
	\item We start by selecting $i\in \mD$ and $j\in \mY$;
	\item Then we extend $\mD$ with a minimum number of node signals that satisfies the parallel path and loop Condition \ref{propx1}.
	\item Every node signal $w_k$ in $w_{\smD}$ for which there is a \emph{direct} or \emph{indirect} confounding variable for the estimation problem $w_{k} \rightarrow w_{\smY}$ is included in $\mY$ and $\mQ$.
	\item With $\mA := \mD \backslash \mQ$ and $\mB = \emptyset$ it follows that by construction there are no confounding variables for the estimation problem $w_{\smA} \rightarrow w_{\smY}$.
	\item Finally, we define the identification setup as the estimation problem $w_{\mathcal D} \rightarrow w_{\mathcal Y}$, with $\mD = \mQ \cup \mA$.
\end{itemize}

As we can observe, the algorithm does not require selection of set $\mB$. This is attributed to the way we handle the indirect confounding variables for the estimation problem $w_{\smA} \rightarrow w_{\smY}$. Instead of tackling these confounding variables by adding blocking node signals $w_{\smB}$ (as in full input case) to be added as predictor inputs, we deal with them by moving the concerned $w_{k}, k \in \mA$ to $w_{\smQ}$ and thus to the set of predicted outputs. We choose this approach in order to minimize the required number of measured node signals. In this way, by construction, there will be no {\it direct} or {\it indirect} confounding variables for the estimation problem $w_{\smA} \rightarrow w_{\smY}$. From this result, we can guarantee that the conditions in Theorem \ref{theomodinv} will be satisfied since $\mB = \emptyset$. Thus at the end of the algorithm we obtain a set of signals that provides target module invariance.

\medskip
\begin{example}\label{exam5}
	Consider the same network as in example \ref{exam4} represented by Figure \ref{fig3}. Applying the algorithm of this section, we first include the input of the target module $w_2$ in $w_{\smD}$ and the output of the target module $w_1$ in $w_{\smY}$. There exist two parallel paths from $w_2$ to $w_1$, namely $w_2 \rightarrow w_3 \rightarrow w_1$ and $w_2 \rightarrow w_3 \rightarrow w_4  \rightarrow w_1$ and no loops through $w_1$. In order to satisfy Condition \ref{condx1} we can include either $w_3$ in $\mD$ such that $\mD=\{2,3\}$ or both $w_3, w_4$ in $\mD$ such that $\mD=\{2,3,4\}$. We choose the former to have minimum number of node signals. Because of the correlation between $v_2$ and $v_1$ there is a confounding variable for the estimation problem  $w_2 \rightarrow w_1$. According to step 3 of the algorithm, $w_2$ is then moved to $\mY$ and $\mQ$, leading to $w_{\smY} = \{w_1,w_2\}$. Because of this change of $\mY$ we have to recheck for presence of confounding variables. However this change does not introduce any additional confounding variables. The resulting estimation problem is $(w_2,w_3) \rightarrow (w_1,w_2)$ with $w_{\smA} = w_3$, $w_{\smB} = \emptyset$, $w_{\smQ} = w_2$ and $w_{\smY} = (w_1,w_2)$. \hfill $\Box$
\end{example}

In comparison with the full input case, the algorithm in this section will typically have a higher number of predicted output nodes and a smaller number of predictor inputs. This implies that there is a stronger emphasis on estimating a (multivariate) noise model $\bar H$.
Given the choice of the direct identification method, and the choice of signals to satisfy the parallel path and loop condition, this algorithm indeed adds the smallest number of additional signals to be measured, as the removal of any of the additional signals will lead to conflicts with the required conditions.
%
%
\section{Algorithm for signal selection: User selection case}
\label{sec:usersel}
Next we focus on the situation that we have a prior given set of nodes that we have access to i.e. a set of nodes that can (possibly) be measured. We refer to these nodes as \emph{accessible nodes} while the remaining nodes are called \emph{inaccessible}. This strategy is different from the \emph{full input case} since we do not assume that we have access to all in-neighbours of $w_{\smY}$.
This will lead to an identification setup with use of accessible node signals that contain information which is relevant for modeling our target module $G_{ji}$.
We consider the situation that nodes $w_i$ and $w_j$ are accessible nodes and there are accessible nodes that satisfy the parallel path and loop Condition \ref{condx1}.\\
The following strategy will be followed:
\begin{enumerate}
	\item We start by selecting $i\in \mD$ and $j\in \mY$;
	\item Then we extend $\mD$ to satisfy the parallel path and loop Condition \ref{condx1};
	\item We include in $\mD$ all accessible $w$-in-neighbors of $\mY$;
\item We extend $\mD$ in such a way that for every non-accessible $w$-in-neighbor $w_k$ of $w_{\smY}$ we include all accessible nodes that have path to $w_k$ that runs through non-accessible nodes only.
\item If there is a direct confounding variable for $w_i \rightarrow w_{\smY}$, or an indirect one that has a path to $w_i$ that does not pass through any accessible nodes, then $i$ is included in $\mY$ and $\mQ$;
    \item A node signal $w_k$, $k \in \mD$ is included in $\mA$ if there are either no confounding variables for $w_k \rightarrow w_{\smY}$ or only indirect confounding variables that have paths to $w_k$ that pass through accessible nodes.
     \item Every node signal $w_k, k \in\mD\backslash\{i\}$ that has a \emph{direct} confounding variable for $w_{k} \rightarrow w_{\smY}$, or an \emph{indirect} confounding variable with a path to $w_k$ that does not pass through any accessible nodes is:
\begin{itemize}
	\item included in $\mB$ if condition 2a and 2c of Theorem \ref{theomodinv} are satisfied on including it in $w_{\smB}$ (else)
	\item included in $\mY$ and $\mQ$; return to step 3.
\end{itemize}
     \item Every node signal $w_k$, $k\in \mA$ for which there are only indirect confounding variables as meant in Step 6, is
        \begin{itemize}
        \item moved to $\mB$ if Conditions 2a and 2c of Theorem \ref{theomodinv} are satisfied and $k\neq i$; (else)
        \item kept in $\mA$ while a set of accessible nodes that blocks the path of the confounding variable is added to $\mB\cup\mA$, while satisfying Conditions 2a and 2c of Theorem \ref{theomodinv}; (else)
        \item included in $\mY$ and $\mQ$;
        \end{itemize}

	\item By construction there are no confounding variables for $w_{\smA} \rightarrow w_{\smY}$.
\end{enumerate}

\medskip
In the algorithm above, the prime reasoning is to deal with confounding variables for $w_{\smA} \rightarrow w_{\smY}$. Direct confounding variables lead to including the respective node in the outputs $\mY$ or shifting the respective input node to $\mB$, while indirect confounding variables are treated by either shifting the input node to $\mB$ or, if its effect can be blocked, by adding an accessible node to the inputs in $\mB$, or, if the blocking conditions can not be satisfied, by including the node in the output $\mY$. Note that the algorithm always provides a solution if Condition 1 of Theorem \ref{theomodinv} (parallel path and loop condition) can be satisfied.


\begin{figure}[htb]
	\includegraphics[scale=0.5]{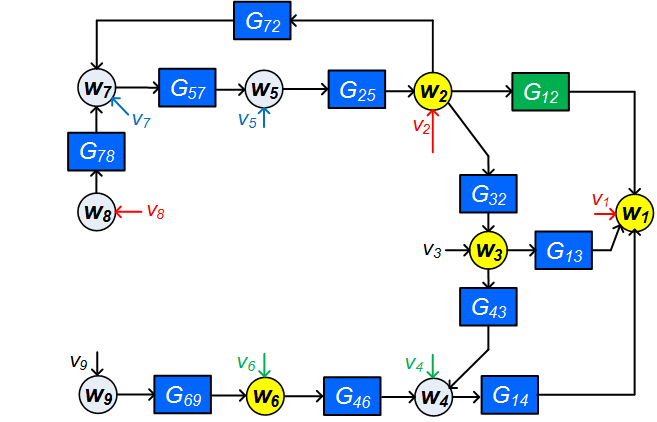}
	\caption{Example network of Figure \ref{fig3} with accessible nodes $w_1$, $w_2$, $w_3$, $w_6$ indicated in yellow.}
	\label{fig4}
\end{figure}

\begin{example}
	Consider the same network as in example \ref{exam4} represented by Figure \ref{fig4}. However, we are given that only the nodes $w_1, w_2, w_3$ and $w_6$ are accessible. We now select the signals according to the algorithm presented in this section. First we include $w_i = w_2$ in $w_{\smD}$ and $w_j = w_1$ in $w_{\smY}$. Then we extend $\mD$ such that the parallel path and loop Condition \ref{condx1} is satisfied. This is done by selecting $\mD = \{2,3\}$. According to step 4, we extend $\mD$ by node $w_6$ as it serves as nearest accessible in-neighbor of $w_4$, being an inaccessible in-neighbor of $w_1$.
As per Step 5, since $v_1$ and $v_2$ are correlated, $w_2$ is moved to $\mY$ and $\mQ$.
As per Step 6, there are no confounding variables for the estimation problem $w_3 \rightarrow w_1$ and hence $w_3$ is included in $w_{\smA}$. Since $v_4$ and $v_6$ are correlated, it implies that there is an \emph{indirect} confounding variable for the estimation problem $w_{6} \rightarrow w_1$, which however does not pass through an accessible node.
Step 7 does not apply since $w_3 \in w_{\smA}$ has no confounding variables.
Step 8 requires to deal with the indirect confounding variable $v_4$ for $w_6 \rightarrow w_1$. Checking Conditions 2a and 2c of Theorem \ref{theomodinv} for $\mA$ and $\mB$, it appears that every path from $w_i = w_2$ or from $w_j = w_1$ to $w_6$ passes through a measured node and there are no confounding variable for the estimation problem $w_{\smA} \rightarrow w_{6}$. Hence we include $w_6$ in $w_{\smB}$.
As a result, the estimation problem is $(w_2,w_3,w_6) \rightarrow (w_1,w_2)$.
\end{example}

\begin{remark}
Rather than starting the signal selection problem from a fixed set of accessible notes, the provided theory allows for an iterative and interactive algorithm for selecting accessible nodes in sensor allocation problems in a flexible way.
%
\end{remark}

\section{Discussion}
All three presented algorithms lead to a set of selected node signals that satisfy the conditions for target module invariance, and thus provide a predictor model in which no confounding variables can deteriorate the estimation of the target module.
Only in the ``User selection case'' this is conditioned on the fact that appropriate node signals should be available to satisfy the parallel path and loop condition. Under these circumstances the presented algorithms are sound and complete \cite{Kroening&Strichmann:16}. This attractive feasibility result is mainly attributed to the addition of predicted outputs, that adds flexibility to solve the problem of confounding variables.

\begin{figure}[htb]
\centering
	\includegraphics[scale=0.65]{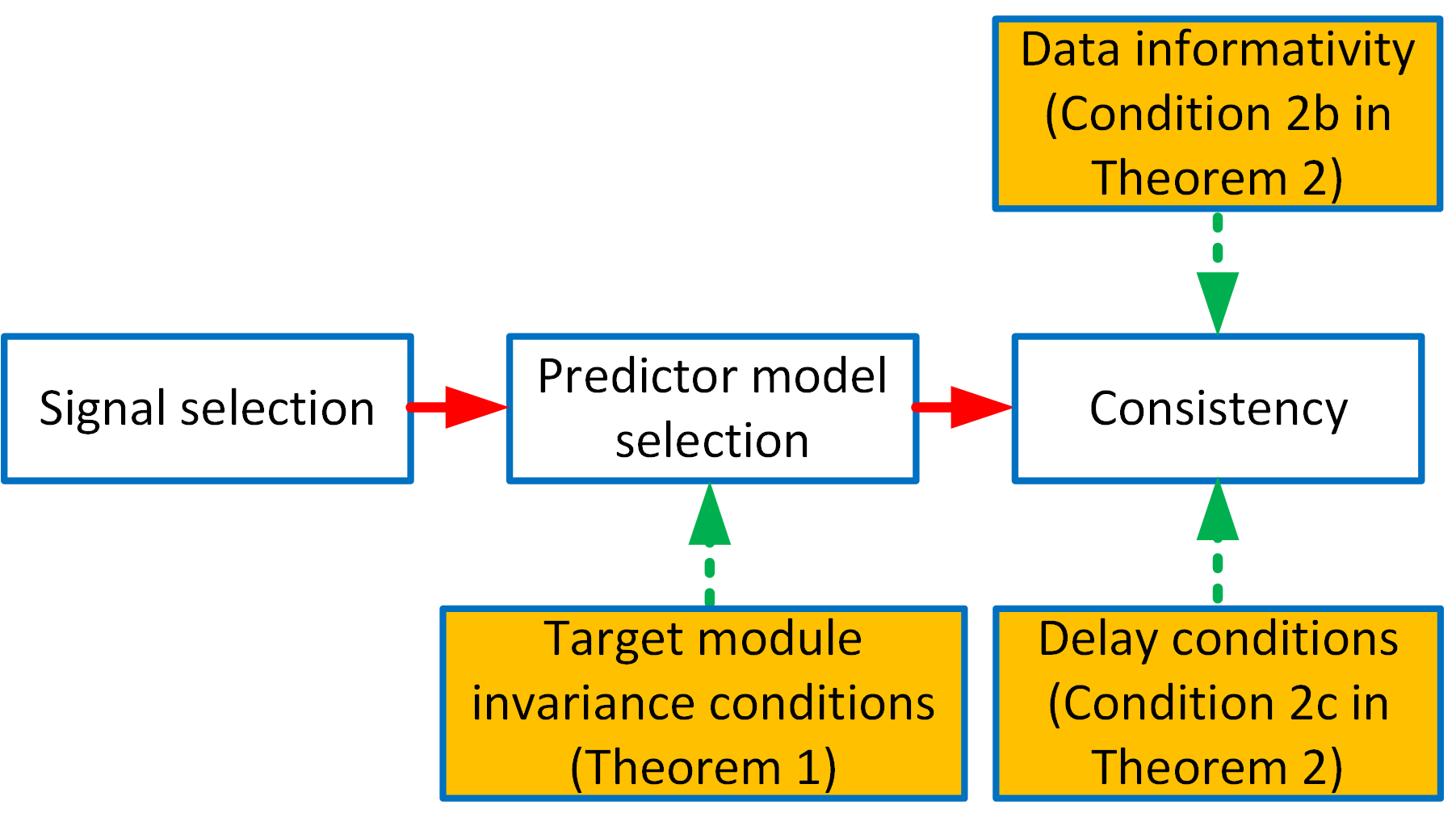}
	\caption{Figure to depict that consistency result requires satisfaction of conditions in Theorem \ref{theorem1} along with the appropriate predictor model.}
	\label{fig8}
\end{figure}

Note that the presented algorithms do not guarantee the consistency of the estimated target module. For this to hold the additional conditions for consistency, among which data-informativity and the delay in path/loop condition, need to be satisfied too, as illustrated in Figure \ref{fig8}. A specification of path-based conditions for data-informativity is beyond the scope of this paper, but first results on this problem are presented in \cite{VandenHof&Ramaswamy_CDC:20}. Including these path-based conditions in the signal selection algorithms would be a next natural step to take. This also holds for the development of data-driven techniques to estimate the correlation structure of the disturbances.

It can be observed that the three algorithms presented in the previous sections rely only on the graphical conditions of the network. This paves way to automate the signal selection procedure using graph based algorithms that are scalable to large dimensions, with input being topology of the network and disturbance correlation structure represented as adjacency matrices. Also,
it can be observed that the three considered cases in the previous sections, most likely will lead to three different experimental setups for estimating the single target module. For all three cases we can arrive at consistent and maximum likehood estimates of the target module. However, because of the fact that the experimental setups are different in the three cases, the data-informativity conditions and the statistical properties of the target module estimates will be different. The minimum variance expressions, in the form of the related Cram\'er-Rao lower bounds, will typically be different for the different experimental setups. Comparing these bounds for different experimental setups is beyond the scope of the current paper and considered as topic for future research.\\
We have formulated identification criteria in the realm of classical prediction error methods. This will typically lead to complex non-convex optimization problems that will scale poorly with the dimensions (number of parameters) of the problems. However alternative optimization approaches are becoming available that scale well and that rely on regularized kernel-based methods, thus exploiting new developments that originate from machine learning, see e.g. \cite{Ramaswamy&etal_CDC:18}, and relaxations that rely on sequential convex optimization, see e.g. \cite{Weerts&etal_IFAC:18,Galrinho&etal_TAC:19}.

\section{CONCLUSIONS}
A new local module identification approach has been presented to identify local modules in a dynamic network with given topology and process noise that is correlated over the different nodes. For this case, it is shown that the problem can be solved by moving from a MISO to a MIMO identification setup. In this setup the target module is embedded in a MIMO problem with appropriately chosen inputs and outputs, that warrant the consistent estimation of the target module with maximum likelihood properties. The key part of the procedure is the handling of direct and indirect confounding variables that are induced by correlated disturbances and/or non-measured node signals, and thus essentially dependent on the (Boolean) topology of the network and the (Boolean) correlation structure of the disturbances. A general theory has been developed that allows for specification of different types of algorithms, of which the ``full input case'', the ``minimum input case'' and the ``user selection case'' have been illustrated through examples. The presented theory is suitable for generalization to the estimation of sets of target modules.

\appendices
\section{Proof of Proposition \ref{propx1}}
Starting with the network representation \eqref{eq1}, we can eliminate the non-measured node variables $w_{\smZ}$ from the equations, by writing the last (block) row of \eqref{eq1} into an explicit expression for $w_{\smZ}$:
\[
w_{\smZ} = (I-G_{\smZ\smZ})^{-1}\left[ \sum_{k \in \mQ \cup \{o\} \cup \mU}\!\!\!\! G_{\smZ k}w_{k} + \sum_{\ell \in \mQ \cup \{o\} \cup \mU \cup \mZ}\!\!\!\! H_{\smZ\ell}w_{\ell} \right],
\]
and by substituting this $w_{\smZ}$ into the expressions for the remaining $w$-variables. As a result
\beqr
	\label{eq13xab}
	\begin{bmatrix} w_{\smQ} \\ w_o \\ w_{\smU}  \end{bmatrix} =
	\begin{bmatrix} \breve G_{\smQ\smQ} & \breve G_{\smQ o} & \breve G_{\smQ\smU}  \\
		\breve G_{o\smQ} & \breve G_{oo} & \breve G_{o\smU} \\
		\breve G_{\smU\smQ} & \breve G_{\smU o} & \breve G_{\smU\smU}  \end{bmatrix}
	\begin{bmatrix} w_{\smQ} \\ w_o \\ w_{\smU} \end{bmatrix} +
	\breve v, \nonumber \\
	\breve v = \breve H\begin{bmatrix} e_{\smQ} \\ e_o \\ e_{\smU} \\ e_{\smZ} \end{bmatrix} = \begin{bmatrix} \breve H_{\smQ\smQ} & \breve H_{\smQ o} & \breve H_{\smQ\smU} & \breve H_{\smQ\smZ} \\
		\breve H_{o\smQ} & \breve H_{oo} & \breve H_{o\smU} & \breve H_{o\smZ}  \\
		\breve H_{\smU\smQ} & \breve H_{\smU o} & \breve H_{\smU\smU} & \breve H_{\smU\smZ} \end{bmatrix}\!\!\!
	\begin{bmatrix} e_{\smQ} \\ e_o \\ e_{\smU} \\ e_{\smZ} \end{bmatrix}
	\eeqr
	with $cov(e)=I$, and where
\begin{equation}
\label{breveG}
\breve G_{kh} = G_{kh} + G_{k\smZ}(I-G_{\smZ\smZ})^{-1}G_{\smZ h}
\end{equation}
with $k,h \in \{\mQ \cup \{o\} \cup \mU\}$, and
\begin{equation}
\label{eqhos1ba}
\breve H_{k\ell} = H_{k\ell} + G_{k\smZ}(I-G_{\smZ\smZ})^{-1}H_{\smZ \ell},
\end{equation}
with $\ell \in \{\mQ \cup \{o\} \cup \mU \cup \mZ\}$.

On the basis of (\ref{eq13xab}), the spectral density of $\breve v$ is given by $\Phi_{\breve v}= \breve H\breve H^*$. Applying a spectral factorization \cite{Youla:61} to $\Phi_{\breve v}$ will deliver $\Phi_{\breve v} =
\tilde H \tilde\Lambda \tilde H^*$ with $\tilde H$ a monic, stable and minimum phase rational matrix, and $\tilde\Lambda$ a positive definite (constant) matrix. Then there exists a white noise process $\tilde\xi$ defined by $\tilde\xi := \tilde H^{-1}\breve H e$ such that $\tilde H \tilde\xi = \breve v$, with cov($\tilde\xi$) = $\tilde\Lambda$, while $\tilde H$ is of the form
\beq
	\label{eqspec}
	\tilde H = \begin{bmatrix}
		\tilde{H}_{11} & \tilde{H}_{12} & \tilde{H}_{13}\\
		\tilde{H}_{21} & \tilde{H}_{22} & \tilde{H}_{23}\\
		\tilde{H}_{31} & \tilde{H}_{32} & \tilde{H}_{33}\\
	\end{bmatrix}
\eeq
and where the block dimensions are conformable to the dimensions of $w_{\smQ}$, $w_o$ and $w_{\smU}$ respectively.
As a result, \eqref{eq13xab} can be rewritten as
\beq
\label{eqims}
\begin{bmatrix} w_{\smQ} \\ w_o \\ w_{\smU}  \end{bmatrix} =
	\begin{bmatrix} \breve G_{\smQ\smQ} & \breve G_{\smQ o} & \breve G_{\smQ\smU}  \\
		\breve G_{o\smQ} & \breve G_{oo} & \breve G_{o\smU} \\
		\breve G_{\smU\smQ} & \breve G_{\smU o} & \breve G_{\smU\smU}  \end{bmatrix}
	\begin{bmatrix} w_{\smQ} \\ w_o \\ w_{\smU} \end{bmatrix} + \tilde H
\begin{bmatrix} \tilde\xi_{\smQ} \\ \tilde\xi_o \\ \tilde\xi_{\smU} \end{bmatrix}.
\eeq
By denoting
\beq \label{Hcheck}
\begin{bmatrix}
\check{H}_{13}\\
\check{H}_{23}
	\end{bmatrix} :=
\begin{bmatrix}
\tilde{H}_{13}\tilde H_{33}^{-1} \\
\tilde{H}_{23}\tilde H_{33}^{-1}
	\end{bmatrix}
\eeq
and premultiplying \eqref{eqims} with
\beq
\begin{bmatrix}
I & 0 & -\check{H}_{13}\\
0 & I & -\check{H}_{23}\\
0 & 0 & I
\end{bmatrix}
\eeq
while only keeping the identity terms on the left hand side, we obtain an equivalent network equation:
\beq
\label{eqims1}
\begin{bmatrix} \!w_{\smQ}\! \\ \!w_o\! \\ \!w_{\smU}\!  \end{bmatrix} \!\!\!=\!\!\!
\begin{bmatrix} \!\breve G_{\smQ\smQ}' \!\!&\!\! \breve G_{\smQ o}' \!\!&\!\! \breve G_{\smQ\smU}'\!  \\
\!\breve G_{o\smQ}' \!&\! \breve G_{oo}' \!\!&\!\! \breve G_{o\smU}'\! \\
\!\breve G_{\smU\smQ} \!\!&\!\! \breve G_{\smU o} \!\!&\!\! \breve G_{\smU\smU}\! \end{bmatrix}
\!\!\begin{bmatrix} \!w_{\smQ}\! \\ \!w_o\! \\ \!w_{\smU}\! \end{bmatrix}\!\!
+\!\! \begin{bmatrix}
		\!\tilde{H}_{11}' \!\!&\!\! \tilde{H}_{12}' \!\!&\!\! 0\!\\
		\!\tilde{H}_{21}' \!\!&\!\! \tilde{H}_{22}' \!\!&\!\! 0\!\\
		\!\tilde{H}_{31} \!\!&\!\! \tilde{H}_{32} \!\!&\!\! \tilde{H}_{33}\!
	\end{bmatrix}\!\!
\begin{bmatrix} \!\tilde\xi_{\smQ}\! \\ \!\tilde\xi_o\! \\ \!\tilde\xi_{\smU}\! \end{bmatrix}\!,
\eeq
with
\beqr \label{eqx81}\label{eqbGqid}
\breve G_{\smQ\smU}' & = & \breve G_{\smQ\smU}-\check{H}_{13}\breve G_{\smU\smU} + \check H_{13}\\\label{eqbGqqd}
\breve G_{\smQ\star}' & = & \breve G_{\smQ\star}-\check{H}_{13}\breve G_{\smU\star}\\\label{eqGood}
\breve G_{o\star}' & = & \breve G_{o\star}-\check{H}_{23}\breve G_{\smU\star}\\\label{eqGoid}
\breve G_{o\smU}' & = & \breve G_{o\smU}-\check{H}_{23}\breve G_{\smU\smU} + \check H_{23}\\\label{eq9}
\tilde H_{1\square}' & = & \tilde H_{1\square} - \check H_{13}\tilde H_{3\square} \\
\tilde H_{2\square}' & = & \tilde H_{2\square} - \check H_{23}\tilde H_{3\square} .
\eeqr
where $\star \in \{\mQ \cup \{o\}\}$ and $\square \in \{1,2\}$.

The next step is now to show that that the block elements $\breve G_{\smQ o}'$ and $\breve G_{oo}'$ in $G$ can be made $0$. This can be done by variable substitution as follows:

The second row in (\ref{eqims1}) is replaced by an explicit expression for $w_o$ according to
\[ w_o = (1-\breve G_{oo}')^{-1}[ \breve G_{o\smQ}'w_{\smQ} + \breve G_{o\smU}'w_{\smU}+ \tilde H_{21}'\tilde\xi_{\smQ} + \tilde H_{22}'\tilde\xi_o]. \]

Additionally, this expression for $w_o$ is substituted into the first block row of (\ref{eqims1}), to remove the $w_o$-dependent term on the right hand side, leading to
\beq
\begin{split}
	\begin{bmatrix} \! w_{\smQ} \! \\ \! w_o \! \\ \! w_{\smU} \! \end{bmatrix} \!\!\! = \!\! &
	\begin{bmatrix} \! \breve G_{\smQ\smQ}'' \! & \! 0 \! & \! \breve G_{\smQ\smU}''\!\\
		\! \bar G_{o\smQ} \! & \! 0 \! & \! \bar G_{o\smU} \!\\
		\! \breve G_{\smU\smQ} \! & \! \breve G_{\smU o} \! & \! \breve G_{\smU\smU} \!\end{bmatrix} \!\!\!
	\begin{bmatrix} \! w_{\smQ} \! \\ \! w_o \! \\ \! w_{\smU} \!   \end{bmatrix} \!\!\!
	+ \!\!\! \begin{bmatrix}
		\! \tilde{H}_{11}'' \!\! & \!\! \tilde{H}_{12}'' \!\! & \!\! 0 \!\\
		\! \tilde{H}_{21}'' \!\! & \!\! \tilde{H}_{22}'' \!\! & \!\! 0 \!\\
		\! \tilde{H}_{31} \!\! & \!\! \tilde{H}_{32} \!\! & \!\! \tilde{H}_{33}
	\end{bmatrix}\!\!\!
	\begin{bmatrix}
		\tilde\xi_{\smQ} \\
		\tilde\xi_o \\
		\tilde\xi_{\smU}
	\end{bmatrix}
\end{split}
\eeq
with
\beqr
\bar G_{o\star} & = & (I - \breve G_{oo}')^{-1}\breve G_{o\star}' \label{eqinvar}\\
\tilde H_{2\star}'' & = & (I - \breve G_{oo}')^{-1}\tilde H_{2\star}'  \\
\breve G_{\smQ\star}'' & = & \breve G_{\smQ\star}' + \breve G_{\smQ o}'\bar G_{o\star}\label{eqGqqdd}\\
\tilde H_{1\star}'' & = & \tilde H_{1\star}' + \breve G_{\smQ o}'\tilde H_{2\star}''.
\eeqr

Since because of these operations, the matrix $\breve G_{\smQ\smQ}''$ might not be hollow, we move any diagonal terms of this matrix to the left hand side of the equation, and premultiply the first (block) equation by the diagonal matrix $(I - \mathrm{diag}(\breve G_{\smQ\smQ}''))^{-1}$, to obtain the expression
\beq
\label{eq194xab}
\begin{split}
	\begin{bmatrix} \! w_{\smQ} \! \\ \! w_o \! \\ \! w_{\smU} \! \end{bmatrix} \!\!\! = \!\! &
	\begin{bmatrix} \! \bar G_{\smQ\smQ} \! & \! 0 \! & \! \bar G_{\smQ\smU}\!\\
		\! \bar G_{o\smQ} \! & \! 0 \! & \! \bar G_{o\smU} \!\\
		\! \breve G_{\smU\smQ} \! & \! \breve G_{\smU o} \! & \! \breve G_{\smU\smU} \!\end{bmatrix} \!\!\!
	\begin{bmatrix} \! w_{\smQ} \! \\ \! w_o \! \\ \! w_{\smU} \!   \end{bmatrix} \!\!\!
	+ \!\!\! \begin{bmatrix}
		\! \tilde{H}_{11}''' \!\! & \!\! \tilde{H}_{12}''' \!\! & \!\! 0 \!\\
		\! \tilde{H}_{21}'' \!\! & \!\! \tilde{H}_{22}'' \!\! & \!\! 0 \!\\
		\! \tilde{H}_{31} \!\! & \!\! \tilde{H}_{32} \!\! & \!\! \tilde{H}_{33}
	\end{bmatrix}\!\!\!
	\begin{bmatrix}
		\tilde\xi_{\smQ} \\
		\tilde\xi_o \\
		\tilde\xi_{\smU}
	\end{bmatrix}
\end{split}
\eeq
with
\beqr
\bar G_{\smQ\smQ} & = & (I - \mathrm{diag}(\breve G_{\smQ\smQ}''))^{-1}(\breve G_{\smQ\smQ}'' - \mathrm{diag}(\breve G_{\smQ\smQ}'')), \label{eqGqq}\\
\bar G_{\smQ\smU} & = & (I - \mathrm{diag}(\breve G_{\smQ\smQ}''))^{-1}\breve G_{\smQ\smU}'' \label{eqGqi} \\
\tilde H_{1\star}''' & = & (I - \mathrm{diag}(\breve G_{\smQ\smQ}''))^{-1}\tilde H_{1\star}''.
\eeqr

As final step, we need the matrix
$\tilde H_r := \begin{bmatrix}
		\! \tilde{H}_{11}''' \!\! & \!\! \tilde{H}_{12}'''\\
		\! \tilde{H}_{21}'' \!\! & \!\! \tilde{H}_{22}''
	\end{bmatrix}$ to be monic, stable and minimum phase to obtain the representation as in (\ref{eq5a}).
To that end, we consider the stochastic process $\tilde v_{\smY} := \tilde H_r \tilde \xi_{\smY}$ with
$\tilde \xi_{\smY} := \begin{bmatrix} \tilde \xi_{\smQ}^{\top} & \tilde \xi_o^{\top} \end{bmatrix}^{\top}$. The spectral density of $\tilde v_{\smY}$ is then given by $\Phi_{\tilde v_{\smY}} = \tilde H_r \tilde\Lambda_{\smY} \tilde H_r^{\star}$  with $\tilde\Lambda_{\smY}$ the covariance matrix of $\tilde \xi_{\smY}$, that can be decomposed as $\tilde\Lambda_{\smY} = \tilde\Gamma_r \tilde\Gamma_r^T$.
From spectral factorization \cite{Youla:61} it follows that the spectral factor $\tilde H_r \tilde\Gamma_r$ of $\Phi_{\tilde v_{\smY}}$ satisfies
\beq \label{fact}
\tilde H_r \tilde\Gamma_r = \bar H_s D
\eeq
with $\bar H_s$ a stable and minimum phase rational matrix, and $D$ an ``all pass'' stable rational matrix satisfying $D D^{\star} = I$. \\
The signal $\tilde v_{\smY}$ can then be written as
\[ \tilde v_{\smY} = \tilde H_r \tilde\xi_{\smY} = \bar H_s D \tilde\Gamma_r^{-1} \tilde \xi_{\smY}. \]
By defining $\bar H_s^{\infty} := \lim_{z \rightarrow \infty} \bar H_s$, this can be rewritten as
\[ \tilde v_{\smY} = \tilde H_r \tilde\xi_{\smY} = \underbrace{\bar H_s (\bar H_s^{\infty})^{-1}}_{\bar H}\underbrace{\bar H_s^{\infty} D \tilde\Gamma_r^{-1} \tilde \xi_{\smY}}_{\xi_{\smY}}. \]

As a result, $\bar H$ is a monic stable and stably invertible rational matrix, and $\xi_{\smY}$ is a white noise process with spectral density given by $\bar H_s^{\infty}D \tilde\Gamma_r^{-1} \Phi_{\tilde\xi_{\smY}}\tilde\Gamma_r^{-T} D^{\star}(\bar H_s^{\infty})^{T} = \bar H_s^{\infty}(\bar H_s^{\infty})^{T}$. Therefore we can write \eqref{eq194xab} as,
\beq
\label{eq194xab1}
\begin{split}
	\begin{bmatrix} \! w_{\smQ} \! \\ \! w_o \! \\ \! w_{\smU} \! \end{bmatrix} \!\!\! = \!\! &
	\begin{bmatrix} \! \bar G_{\smQ\smQ} \! & \! 0 \! & \! \bar G_{\smQ\smU}\!\\
		\! \bar G_{o\smQ} \! & \! 0 \! & \! \bar G_{o\smU} \!\\
		\! \breve G_{\smU\smQ} \! & \! \breve G_{\smU o} \! & \! \breve G_{\smU\smU} \!\end{bmatrix} \!\!\!
	\begin{bmatrix} \! w_{\smQ} \! \\ \! w_o \! \\ \! w_{\smU} \!   \end{bmatrix} \!\!\!
	+ \!\!\! \begin{bmatrix}
		\! \bar{H}_{11} \!\! & \!\! \bar{H}_{12} \!\! & \!\! 0 \!\\
		\! \bar{H}_{21} \!\! & \!\! \bar{H}_{22} \!\! & \!\! 0 \!\\
		\! \bar{H}_{31} \!\! & \!\! \bar{H}_{32} \!\! & \!\! \tilde{H}_{33}
	\end{bmatrix}\!\!\!
	\begin{bmatrix}
		\xi_{\smQ} \\
		\xi_o \\
		\tilde \xi_{\smU}
	\end{bmatrix}
\end{split}
\eeq
where $\begin{bmatrix}
         \bar H_{31} & \bar H_{32}
       \end{bmatrix} = \begin{bmatrix}
         \tilde H_{31} & \tilde H_{32}
       \end{bmatrix} \tilde\Gamma_r D^{-1} (\bar H_{s}^{\infty})^{-1}$.
Let $[\begin{matrix}\bar H_{31}' & \bar H_{32}'
                           \end{matrix}] = [\begin{matrix}
                             \bar H_{31} & \bar H_{32}
                           \end{matrix}]\begin{bmatrix}
                \bar H_{11} & \bar H_{12} \\
                \bar H_{21} & \bar H_{22}
              \end{bmatrix}^{-1}$.
Pre-multiplying \eqref{eq194xab1} with $\begin{bmatrix}
I & 0 & 0\\
0 & I & 0\\
-\bar H_{31}' & -\bar H_{32}' & I
\end{bmatrix}$ while only keeping the identity terms on the left hand side, we obtain an equivalent network equation:
\beq
\label{eq194xab2}
\begin{split}
	\begin{bmatrix} \! w_{\smQ} \! \\ \! w_o \! \\ \! w_{\smU} \! \end{bmatrix} \!\!\! = \!\! &
	\begin{bmatrix} \! \bar G_{\smQ\smQ} \! & \! 0 \! & \! \bar G_{\smQ\smU}\!\\
		\! \bar G_{o\smQ} \! & \! 0 \! & \! \bar G_{o\smU} \!\\
		\! \breve G_{\smU\smQ}' \! & \! \breve G_{\smU o}' \! & \! \breve G_{\smU\smU}' \!\end{bmatrix} \!\!\!
	\begin{bmatrix} \! w_{\smQ} \! \\ \! w_o \! \\ \! w_{\smU} \!   \end{bmatrix} \!\!\!
	+ \!\!\! \begin{bmatrix}
		\! \bar{H}_{11} \!\! & \!\! \bar{H}_{12} \!\! & \!\! 0 \!\\
		\! \bar{H}_{21} \!\! & \!\! \bar{H}_{22} \!\! & \!\! 0 \!\\
		\! 0 \!\! & \!\! 0 \!\! & \!\! \tilde{H}_{33}
	\end{bmatrix}\!\!\!
	\begin{bmatrix}
		\xi_{\smQ} \\
		\xi_o \\
		\tilde \xi_{\smU}
	\end{bmatrix}
\end{split}
\eeq
where $\breve G_{\smU\smQ}'  = \breve G_{\smU\smQ} -\bar H_{31}'\breve G_{\smQ\smQ}''' -\breve H_{32}'\bar G_{o\smQ}'' + \bar H_{31}'$, $\breve G_{\smU o}'  = \breve G_{\smU o} + \bar H_{32}'$ and $\breve G_{\smU\smU}'  = \breve G_{\smU\smU} -\bar H_{31}'\breve G_{\smQ\smU}''' -\bar H_{32}'\breve G_{o\smU}''$.
In order to make $\breve G_{\smU\smU}'$ hollow, we move any diagonal terms of this matrix to the left hand side of the equation, and pre-multiply the third (block) equation by the diagonal matrix $(I - \mathrm{diag}(\breve G_{\smU\smU}'))^{-1}$. This will modify (3,3) (block) element of the $H$ matrix to $(I - \mathrm{diag}(\breve G_{\smU\smU}'))^{-1}\tilde H_{33}$, which we need to be monic, stable and stably invertible. Applying spectral factorization as before \cite{Youla:61}, we can write the term $(I - \mathrm{diag}(\breve G_{\smU\smU}'))^{-1}\tilde H_{33}\tilde \xi_{\smU}$ as $\bar H_{33}\xi_{\smU}$ where $\bar H_{33}$ is monic, stable and stably invertible and $\xi_{\smU}$ is a white noise process with covariance $\Lambda_{33}$.
This completes the proof for obtaining (\ref{eq5a}).

\medskip
The absence of confounding variables for the estimation problem $w_{\smU} \rightarrow w_{\smY}$ can be proved as follows. Since all non-measured nodes $w_{\smZ}$ are removed in the network represented by (\ref{eq5a}), the only non-measured signals in the network are the noise signals in $\xi_m$ and they do not have any unmeasured paths to any nodes in the network (i.e. to $w_m$). Due to the block-diagonal structure of $\bar H_m$ in (\ref{eq5a}), the only non-measured signals that have direct paths to $w_{\smU}$ originate from $\xi_{\smU}$, while the only non-measured signals that have direct paths to $w_{\smY}$ originate from $[\xi_{\smQ}^T\ \xi_o]^T$. Therefore there does not exist an element of $\xi_m$ that has simultaneous unmeasured paths or direct paths to both $w_{\smU}$ and $w_{\smY}$.
\hfill $\Box$

\section{Proof of Theorem \ref{theomodinv}}
In order to prove Theorem \ref{theomodinv}, we first present three preparatory Lemmas.

\begin{lemma}
\label{lemmax1}
Consider a dynamic network as defined in (\ref{eq1}), a vector $e_{\smX}$ of white noise sources  with $\mX \subseteq \mL$, and two subsets of nodes $w_{\Phi}$ and $w_{\Omega}$, $\Phi, \Omega \subset \mL\backslash \mZ$. If in $e_{\smX}$ there is no confounding variable for the estimation problem $w_{\Phi} \rightarrow w_{\Omega}$, then
\[ \breve H_{\Omega \smX} \breve H_{\Phi \smX}^* = \breve H_{\Phi \smX} \breve H_{\Omega \smX}^* = 0, \]
where $\breve H_{\Omega \smX}$, $\breve H_{\Phi \smX}$ are the noise model transfer functions in the immersed network (\ref{eq13xab}) related to the appropriate variables.
\end{lemma}

{\bf Proof:}
If in $e_{\smX}$ there is no confounding variable for the formulated estimation problem, then for all $e_x$, $x \in \mX$ there do not exist simultaneous paths from $e_x$  to $w_{\Phi}$ and $w_\Omega$, that are direct or pass through nodes in $\mZ$ only. \\
For the network where signals $w_{\smZ}$ are immersed, it follows from \eqref{eqhos1ba}, that $\breve H_{k\ell} = H_{k\ell} + G_{k\smZ}(I-G_{\smZ\smZ})^{-1}H_{\smZ \ell}$ where $k \in \Phi$ and $\ell \in \mX$. The first term in the sum (i.e. $H_{kl}$) is the noise model transfer in the direct path from $e_\ell$ to $w_k$ and the second part of the sum is the transfer function in the unmeasured paths (i.e. paths through $w_{\smZ}$ only) from $e_\ell$ to $w_k$. If all paths from a node signal $e_x$ to $w_{\Phi}$ pass through a node in $w_{\mL\backslash \mZ}$, then there are no direct or unmeasured paths from $e_x$ to nodes in $w_{\Phi}$. This implies that $\breve H_{kx} = \breve H_{kx}^{*} = 0$ for all $k \in \Phi$ (i.e $\breve H_{\Phi x} = 0$). A dual reasoning applies to paths from $e_x$ to $w_{\Omega}$.
Consider $e_{\smX} = [\begin{matrix}
              e_{x_1} & e_{x_2} & \dots & e_{x_n}
            \end{matrix}]^{\top}$.
Then we have $\breve H_{\Phi \smX} \breve H_{\Omega \smX}^* = \breve H_{\Phi x_1} \breve H_{\Omega x_1}^* + \dots + \breve H_{\Phi x_n} \breve H_{\Omega x_n}^*$. If the condition in the lemma is satisfied, implying that there do not exist simultaneous paths, then in each of the product terms we either have $\breve H_{\Phi x_k} = 0$ or $\breve H_{\Omega x_k}^* = 0$ where $k = \{1, 2, \dots, n\}$. This proves the result of lemma \ref{lemmax1}.
\hfill $\Box$

\medskip
\begin{lemma}
\label{lemx1b}
Consider a dynamic network as defined in (\ref{eq13xab}) with target module $G_{ji}$, where the non-measured node signals $w_{\smZ}$ are immersed, while the node sets $\{o,\mQ,\mU\}$ are chosen according to the specifications in Section \ref{sec:concepts}.\\
Then $\bar G_{ji}$ is given by the following expressions:
\beq\label{invarji1}
\mbox{If }i \in \mQ:\
\bar G_{ji} \!\!=\!\! (I\! -\! \breve G_{jj} + \check{H}_{j3}\breve G_{\smU j})^{-1}(\breve G_{ji}\! -\! \check{H}_{j3}\breve G_{\smU i})
\eeq
\beq\label{invarji2}
\mbox{If }i \in \mU:
\bar G_{ji} \!\!=\!\! (I\! -\! \breve G_{jj}\! +\! \check{H}_{j3}\breve G_{\smU j})^{-1}(\breve G_{ji}\! -\! \check{H}_{j3}\breve G_{\smU i}\! +\! \check{H}_{ji})
\eeq
where $\check H_{j3}$ is the row vector corresponding to the row of node signal $j$ in $\check H_{13}$ (if $j \in \mQ$) or in $\check H_{23}$ (if $j \in o$), and $\check H_{ji}$ is the element corresponding to the column of node signal $i$ in $\check H_{j3}$.
\end{lemma}

{\bf Proof:}
For the target module $G_{ji}$ we have the following cases that can occur:
\begin{enumerate}
  \item $j = o$  and $i \in \mU$. From \eqref{eqinvar} we have $\bar G_{ji} = (I - \breve G_{jj}')^{-1}\breve G_{ji}'$ where $\breve G_{jj}'$ is given by \eqref{eqGood} and $\breve G_{ji}'$ is given by \eqref{eqGoid}. This directly leads to (\ref{invarji2}).
  \item $j = o$  and $i \in \mQ$. From \eqref{eqinvar} we have $\bar G_{ji} = (I - \breve G_{jj}')^{-1}\breve G_{ji}'$ where $\breve G_{jj}'$ and $\breve G_{ji}'$ are given by \eqref{eqGood}, leading to (\ref{invarji1}).
  \item $j \in \mQ$, $o$ is void and $i \in \mU$. From \eqref{eqGqi} we have $\bar G_{ji} = (I - \breve G_{jj}'')^{-1}\breve G_{ji}''$ where $\breve G_{jj}''$ and $\breve G_{ji}''$ are given by \eqref{eqGqqdd}. Since $o$ is void, \eqref{eqGqqdd} leads to $G_{\smQ\star}'' =  \breve G_{\smQ\star}'$. Therefore $\breve G_{jj}'' = \breve G_{jj}'$ which is specified by \eqref{eqbGqqd}, and $\breve G_{ji}'' = \breve G_{ji}'$ which is given by \eqref{eqbGqid}. This leads to (\ref{invarji2}).
  \item $j \in \mQ$, $o$ is void and $i \in \mQ$. Since $j \neq i$ it follows from \eqref{eqGqq} that
  $\bar G_{ji} = (I - \breve G_{jj}'')^{-1}\breve G_{ji}''$ where $\breve G_{jj}''$ and $\breve G_{ji}''$ are given by \eqref{eqGqqdd}. Since $o$ is void, \eqref{eqGqqdd} leads to $G_{\smQ\star}'' =  \breve G_{\smQ\star}'$. Therefore for this case, $\breve G_{jj}'' = \breve G_{jj}'$ and $\breve G_{ji}'' = \breve G_{ji}'$, which are given by \eqref{eqbGqqd}. This leads to (\ref{invarji1}). 
\end{enumerate}

\medskip
\begin{lemma}
	\label{lemx12}
	Consider a dynamic network as defined in (\ref{eq13xab}) where the non-measured node signals $w_{\smZ}$ are immersed, and let $\mU$ be decomposed in sets $\mA$ and $\mB$ satisfying Condition \ref{condx2a}. Then the spectral density $\Phi_{\breve{v}}$ has the unique spectral factorization
	$\Phi_{\breve v} = \tilde H\Lambda\tilde H^*$ with $\Lambda$ constant and $\tilde H$ monic, stable, minimum phase, and of the form
	\beq
	\label{eq20}
 \Lambda \!\!=\!\! \begin{bmatrix}
		\Lambda_{11} \!&\! \Lambda_{12} \!&\! \Lambda_{13} & 0\\
		\Lambda_{21} \!&\! \Lambda_{22} \!&\! \Lambda_{23} & 0\\
		\Lambda_{31} \!&\! \Lambda_{32} \!&\! \Lambda_{33} & 0\\
		0 \!&\! 0 \!&\! 0 \!&\! \Lambda_{44}\\
	\end{bmatrix}, \!\!\ \
	\tilde H \!\!=\!\! \begin{bmatrix}
		\tilde{H}_{11} \!&\! \tilde{H}_{12} \!&\! \tilde{H}_{\smQ\smB} \!&\! 0\\
		\tilde{H}_{21} \!&\! \tilde{H}_{22} \!&\! \tilde{H}_{o\smB} \!&\! 0\\
		\tilde{H}_{\smB\smQ} \!&\! \tilde{H}_{\smB o} \!&\! \tilde{H}_{\smB\smB} \!&\! 0\\
		0 \!&\! 0 \!&\! 0 \!&\! \tilde{H}_{\smA\smA}\\
	\end{bmatrix}, \
	\eeq
	where the block dimensions are conformable to the dimensions of $w_{\smQ}$, $w_o$, $w_{\smB}$ and $w_{\smA}$ respectively.
\end{lemma}

{\bf Proof:}
On the basis of (\ref{eq13xab}) we write $w_{\smU} = [\begin{matrix}
                       w_{\smB}^{\top} & w_{\smA}^{\top}
                     \end{matrix}]^{\top}$ and
                     \beq \label{HBsplit}
                     \breve v = \breve H\begin{bmatrix} e_{\smQ} \\ e_o \\ e_{\smB} \\e_{\smA} \\ e_{\smZ} \end{bmatrix} = \begin{bmatrix} \breve H_{\smQ\smQ} & \breve H_{\smQ o} & \breve H_{\smQ\smB} & \breve H_{\smQ\smA} & \breve H_{\smQ\smZ} \\
		\breve H_{o\smQ} & \breve H_{oo} & \breve H_{o\smB} & \breve H_{o\smA} & \breve H_{o\smZ}  \\
		\breve H_{\smB\smQ} & \breve H_{\smB o} & \breve H_{\smB\smB} & \breve H_{\smB\smA} & \breve H_{\smB\smZ}  \\
		\breve H_{\smA\smQ} & \breve H_{\smA o} & \breve H_{\smA\smB} & \breve H_{\smA\smA} & \breve H_{\smA\smZ} \end{bmatrix}\!\!\!
	\begin{bmatrix} e_{\smQ} \\ e_o \\ e_{\smB} \\e_{\smA} \\ e_{\smZ} \end{bmatrix}
                     \eeq
with $cov(e) = I$ and the components of $\breve H$ as specified in (\ref{eqhos1ba}).
Starting from the expression (\ref{HBsplit}), the spectral density $\Phi_{\breve v}$ can be written as $\breve H\breve H^*$ while it is denoted as
\beq\label{eq32xx}
\Phi_{\breve v} = \begin{bmatrix} \Phi_{\breve v_{\smQ}} & \Phi_{\breve v_{\smQ}\breve v_o} & \Phi_{\breve v_{\smQ}\breve v_{\smB}} & \Phi_{\breve v_{\smQ}\breve v_{\smA}}\\
	\Phi_{\breve v_{\smQ}\breve v_o}^* & \Phi_{\breve v_o} & \Phi_{\breve v_o\breve v_{\smB}} & \Phi_{\breve v_o\breve v_{\smA}} \\
	\Phi_{\breve v_{\smQ}\breve v_{\smB}}^* & \Phi_{\breve v_o\breve v_{\smB}}^* & \Phi_{\breve v_{\smB}} & \Phi_{\breve v_{\smB}\breve v_{\smA}} \\
	\Phi_{\breve v_{\smQ}\breve v_{\smA}}^* & \Phi_{\breve v_o\breve v_{\smA}}^* & \Phi_{\breve v_{\smB}\breve v_{\smA}}^* & \Phi_{\breve v_{\smA}} \end{bmatrix}.
\eeq
In this structure we are particularly going to analyse the elements
\beq
\resizebox{1.0\hsize}{!}{$
	\begin{split}
	\Phi_{\breve v_{\smQ}\breve v_{\smA}} = & \breve H_{\smQ\smQ}\breve H_{\smA\smQ}^* + \breve H_{\smQ o}\breve H_{\smA o}^* + \breve H_{\smQ\smB}\breve H_{\smA\smB}^*+ \breve H_{\smQ\smA}\breve H_{\smA\smA}^*+ \breve H_{\smQ\smZ}\breve H_{\smA\smZ}^* \\
	\Phi_{\breve v_o\breve v_{\smA}} = & \breve H_{o\smQ}\breve H_{\smA\smQ}^* + \breve H_{oo}\breve H_{\smA o}^* + \breve H_{o\smB}\breve H_{\smA\smB}^*+ \breve H_{o\smA}\breve H_{\smA\smA}^* +\breve H_{o\smZ}\breve H_{\smA\smZ}^*\\
	\Phi_{\breve v_{\smB}\breve v_{\smA}} = & \breve H_{\smB\smQ}\breve H_{\smA\smQ}^* + \breve H_{\smB o}\breve H_{\smA o}^* + \breve H_{\smB\smB}\breve H_{\smA\smB}^* + \breve H_{\smB\smA}\breve H_{\smA\smA}^* + \breve H_{\smB\smZ}\breve H_{\smA\smZ}^* \label{eqfxbi}
	\end{split}
$}
\eeq
If $\mA$ and $\mB$ satisfy Condition \ref{condx2a}, then none of the white noise terms $e_x$, $x \in \mL$ will be a confounding variable for the estimation problems $w_{\smA} \rightarrow w_{\smQ}$, $w_{\smA} \rightarrow w_{o}$ or $w_{\smA} \rightarrow w_{\smB}$. Then it follows from Lemma \ref{lemmax1} that all of the terms in (\ref{eqfxbi}) are zero.
As a result we can write the spectrum in equation \eqref{eq32xx} as,
\beq\label{eq34}
\Phi_{\breve v} = \begin{bmatrix} \Phi_{\breve v_{\smQ}} & \Phi_{\breve v_{\smQ}\breve v_o} & \Phi_{\breve v_{\smQ}\breve v_{\smB}} & 0\\
	\Phi_{\breve v_{\smQ}\breve v_o}^* & \Phi_{\breve v_o} & \Phi_{\breve v_o\breve v_{\smB}} & 0 \\
	\Phi_{\breve v_{\smQ}\breve v_{\smB}}^* & \Phi_{\breve v_o\breve v_{\smB}}^* & \Phi_{\breve v_{\smB}} & 0 \\
	0 & 0 & 0 & \Phi_{\breve v_{\smA}} \end{bmatrix}
\eeq
Then the spectral density $\Phi_{\breve{v}}$ has the unique spectral factorization \cite{Youla:61}
\beq
\Phi_{\breve v} = \begin{bmatrix}
F_{11}\Lambda_{1}F_{11}^* & 0 \\ 0 & F_{22}\Lambda_{2}F_{22}^*
\end{bmatrix}= \tilde H\Lambda\tilde H^*
\eeq
where $\tilde H$ is of the form in (\ref{eq20}), and monic, stable and minimum phase. \hfill $\Box$

\medskip
Next we proceed with the proof of Theorem \ref{theomodinv}.

With Lemma \ref{lemx1b} it follows that $\bar G_{ji}$ is given by either \eqref{invarji1} or \eqref{invarji2}. For analysing these two expressions, we first are going to specify $\breve G_{ji}$ and $\breve G_{jj}$. From \eqref{breveG}, we have
\beqr
\breve G_{ji} & = & G_{ji} + G_{j\smZ}(I - G_{\smZ\smZ})^{-1}G_{\smZ i} \label{eq100}\\
\breve G_{jj} & = & G_{jj} + G_{j\smZ}(I - G_{\smZ\smZ})^{-1}G_{\smZ j}, \label{eq101}
\eeqr
where the first terms on the right hand sides reflect the direct connections from $w_i$ to $w_j$ (respectively from $w_j$ to $w_j$) and the second terms reflect the connections that pass only through nodes in $\mZ$. By definition,  $G_{jj}=0$ since the $G$ matrix in the network in \eqref{eq1} is hollow. Under the parallel path and loop condition \ref{condx1}, the second terms on the right hand sides of (\ref{eq100}), (\ref{eq101}) are zero, so that $\breve G_{ji} = G_{ji}$ and $\breve G_{jj} = 0$.

What remains to be shown is that in \eqref{invarji1} and \eqref{invarji2}, it holds that
\beq
\label{eq102}
\check{H}_{j3}\breve G_{\smU j} = \check{H}_{j3}\breve G_{\smU i} = 0
\eeq
while additionally for $i \in \mU$, it should hold that
\beq
\label{eq103}
\check H_{ji} = 0.
\eeq
With definition \eqref{Hcheck} for $\check H$ and the special structure of $\tilde H_{13}$ and $\tilde H_{23}$ in (\ref{eqspec}) that is implied by the result (\ref{eq20}) of Lemma \ref{lemx12}, we can write
\beq
\label{eq104}
\begin{bmatrix} \check{H}_{13}\\ \check{H}_{23} \end{bmatrix} =
\begin{bmatrix}
\tilde{H}_{\smQ\smB} & 0\\
\tilde{H}_{o\smB} & 0\\
	\end{bmatrix}
\begin{bmatrix}
\tilde{H}_{\smB\smB} & 0\\
0 & \tilde{H}_{\smA\smA}\\
\end{bmatrix}^{-1} =  \begin{bmatrix}
\check{H}_{\smQ\smB} & 0\\
\check{H}_{o\smB} & 0\\
	\end{bmatrix},
\eeq
implying that columns in this matrix related to inputs $k \in \mA$ are zero.\\
In order to satisfy (\ref{eq103}) we need the condition that: if $i \in \mU$ then $i \in \mA$. This is equivalently formulated as $i \in \mQ \cup \mA$ (conditon 2b). \\
In order to satisfy (\ref{eq102}) we note that $\check H_{j3}$ is a row vector, of which the second part (the columns related to signals in $\mA$) is equal to $0$, according to (\ref{eq104}). Consequently, (\ref{eq102}) is satisfied if for every $k \in \mB$ it holds that $\breve G_{kj} = \breve G_{ki} = 0$. On the basis of \eqref{breveG}, this condition is satisfied if for every $w_k \in w_{\smB}$ there do not exist direct or unmeasured paths from $w_i$ to $w_k$ and from $w_j$ to $w_k$ (condition 2c).
\hfill $\Box$

\section{Proof of Theorem \ref{theorem1}}
Expression \eqref{eq5} can be written as
\[ w_{\smY} = \bar G^o
w_{\smD}
+ \bar H^o \xi_{\smY}.\]
Substituting this into the expression for the prediction error \eqref{eqx5}, leads to
\beq
\label{eq201}
\varepsilon(t,\theta) := \bar H(q,\theta)^{-1}\left[\Delta\bar G(q, \theta)
w_{\smD}
+ \Delta\bar H(q, \theta) \xi_{\smY}\right] + \xi_{\smY}
\eeq
where $\Delta \bar G(q,\theta) = \bar G^o - \bar G(q,\theta)$ and $\Delta \bar H(q,\theta) = \bar H^o - \bar H(q,\theta)$. The proof of consistency involves two steps.
\begin{enumerate}
	\item To show that $\E \varepsilon^T(t,\theta)W\varepsilon(t,\theta)$ achieves its minimum for $\Delta \bar G(\theta) = 0$ and $\Delta \bar H(\theta) = 0$,
	\item To show the conditions under which the minimum is unique.
\end{enumerate}
\textit{Step 1:} With Proposition \ref{propx1} it follows that our data generating system can always be written in the form (\ref{eq5a}), such that $w_m = T(q)\xi_m$. We denote $T_{1}$ as the matrix composed of the first and third (block) row of $T$, such that $w_{\smD} = T_1(q)\xi_m$. Substituting this into (\ref{eq201}) gives
\[ \varepsilon(t,\theta) := \bar H(q,\theta)^{-1}\left[\Delta\bar G(q, \theta)T_{1}
+ \begin{bmatrix} \Delta \bar H(\theta) & 0 \end{bmatrix}\right]\xi_m + \xi_{\smY},
\]
where $\xi_m$ is (block) structured as $[\xi_{\smY}^\top\ \xi_{\smU}^\top]^\top$.\\
In order to prove that the minimum of $\Eb \left[\varepsilon^T(t,\theta)W\varepsilon(t,\theta)\right]$ is attained for $\Delta \bar G(\theta) = 0$ and $\Delta \bar H(\theta) = 0$, it is sufficient to show that
\beq\label{eqx421}
\begin{split}
	\left[ \Delta \bar G(\theta) T_{1}(q) + \begin{bmatrix} \Delta \bar H(\theta) & 0 & 0 \end{bmatrix} \right]\xi_m(t) \\
\end{split}
\eeq
is uncorrelated to $\xi_{\smY}(t)$. In order to show this, let $\mF_n = \mU \backslash \mF$, with $\mF$ as defined in the Theorem, while we decompose $\xi_m$ according to $\xi_m = [\begin{matrix} \xi_{\smY}^\top & \xi_{\smF}^\top & \xi_{\smF_{n}}^\top \end{matrix}]^\top$.
Using a similar block-structure notation for $\Delta \bar G$, $T$ and $\Delta \bar H$, (\ref{eqx421}) can then be written as
\beq \label{eqx431}
\resizebox{1.02\hsize}{!}{$
	\begin{split}
	&\left(\Delta \bar G_{\smY\smQ}(\theta)T_{\smQ\smY} + \Delta \bar G_{\smY\smF}(\theta)T_{\smF\smY} + \Delta \bar G_{\smY\smF_n}(\theta)T_{\smF_n\smY} + \Delta \bar H_{\smY\smY}(\theta)\right)\xi_{\smY} + \\
	&\ +\left(\Delta \bar G_{\smY\smQ}(\theta)T_{\smQ\smF} + \Delta \bar G_{\smY\smF}(\theta)T_{\smF\smF} + \Delta \bar G_{\smY\smF_n}(\theta)T_{\smF_n\smF}\right)\xi_{\smF} \\
	&\ + \left(\Delta \bar G_{\smY\smQ}(\theta)T_{\smQ\smF_n} + \Delta \bar G_{\smY\smF}(\theta)T_{\smF\smF_n} + \Delta \bar G_{\smY\smF_n}(\theta)T_{\smF_n\smF_n}\right)\xi_{\smF_n}.
	\end{split}
$}
\eeq
Since, by definition, $\xi_{\smF_n}(t)$ is statically uncorrelated to $\xi_{\smY}(t)$, the $\xi_{\smF_n}$-dependent term in (\ref{eqx431}) cannot create any static correlation with $\xi_{\smY}(t)$. Then it needs to be shown that the $\xi_{\smY}$- and $\xi_{\smF}$-dependent terms in (\ref{eqx431}) all reflect strictly proper filters. i.e. that they all contain at least a delay.\\
$\Delta \bar H(\theta)$ is strictly proper since both $\bar{H}(\theta)$ and $\bar H^o$ are monic. Therefore, $\Delta \bar H_{\smY\smY}(\theta)$ will have at least a delay in each of its transfers.\\
If all paths from $w_{\smY\cup\smF}$ to $w_{\smY}$ in the \emph{transformed network} and in its parameterized model have at least a delay (as per Condition c in the theorem), then all terms $\Delta \bar G_{\smY\smQ}(\theta)$ and $\Delta \bar G_{\smY\smF}(\theta)$ will have a delay.\\
%
We then need to consider the two remaining terms, $\Delta \bar G_{\smY\smF_n}(\theta)T_{\smF_n\smY}$ and $\Delta \bar G_{\smY\smF_n}(\theta)T_{\smF_n\smF}$. From the definition of $\Delta \bar G_{\smY\smF_n}(\theta)$, each of the two terms can be represented as the sum of two terms. $\bar G_{\smY\smF_n}T_{\smF_n\smY}$ and $\bar G_{\smY\smF_n}T_{\smF_n\smF}$ represent paths from $w_{\smY}$ to $w_{\smY}$ and from $w_{\smF}$ to $w_{\smY}$ respectively in the \emph{transformed network}. Whereas, $\bar G_{\smY\smF_n}(\theta)T_{\smF_n\smY}$ and $\bar G_{\smY\smF_n}(\theta)T_{\smF_n\smF}$ is partly induced by the parameterized model and partly by the paths from $w_{\smY}$ to $w_{\smF_n}$ and from $w_{\smF}$ to $w_{\smF_n}$ respectively in the \emph{transformed network}.
According to condition c of the theorem (delay conditions), these transfer functions are strictly proper. This implies that (\ref{eqx431}) is statically uncorrelated to $\xi_{\smY}(t)$. Therefore we have, $\Eb \left[\varepsilon^T(t,\theta)W\varepsilon(t,\theta)\right] = \Eb \left[{||\Delta X(\theta)\xi_m||}_W\right] + \Eb\left[\xi_{\smY}^\top W \xi_{\smY} \right]$ where $\Delta X(\theta) = \bar H(\theta)^{-1}\left[ \Delta \bar G(\theta) T_{1}(q) + \begin{bmatrix} \Delta \bar H(\theta) & 0 & 0 \end{bmatrix} \right]$. As a result, the minimum of $\Eb \left[\varepsilon^T(t,\theta)W\varepsilon(t,\theta)\right]$, which is $\Eb\left[\xi_{\smY}^\top W \xi_{\smY}\right]$, is achieved for $\Delta \bar G(\theta) = 0$ and $\Delta \bar H(\theta) = 0$.

\textit{Step 2:} When the minimum is achieved, we have $\Eb \left[{||\Delta X(\theta)\xi_m||}_W\right]$ to be zero. From \eqref{eq201}, we have
$
\Delta X(\theta)\xi_m = \bar H(q,\theta)^{-1}\left[\begin{bmatrix}
\Delta\bar G(q, \theta) & \Delta\bar H(q, \theta)
\end{bmatrix} \begin{bmatrix}
w_{\smD}^\top & \xi_{\smY}^\top
\end{bmatrix}^\top\right].
$
Using the expression of $\xi_o$ from \eqref{eq5} and substituting it in the expression of $\Delta X(\theta)\xi_m$ we get,
$
\Delta X(\theta)\xi_m = \bar H(q,\theta)^{-1}\left[\begin{bmatrix}
\Delta\bar G(q, \theta) & \Delta\bar H(q, \theta)
\end{bmatrix}J\kappa(t)\right] = \Delta x(\theta)J\kappa(t)$ where,
\[
J\!\! =\!\! \begin{bmatrix}
I & 0 & 0 \\ 0 & I & 0 \\ -(\bar H_{oo})^{-1}\bar G_{o\smD}\! &\! -(\bar H_{oo})^{-1}\bar H_{o\smQ}\! &\!\! (\bar H_{oo})^{-1}\!
\end{bmatrix}\!; \bar G_{o\smD}^{\top} \! =\! \begin{bmatrix}
\bar G^{\top}_{o\smQ} \\ \bar G^{\top}_{o\smU}
\end{bmatrix}\! .
\]
Writing $\Eb \left[{||\Delta X(\theta)\xi_m||}_W\right] = \Eb \left[{||\Delta x(\theta)J\kappa(t)||}_W\right] = 0$ using Parseval's theorem in the frequency domain, we have
\beq
\frac{1}{2\pi}\int_{-\pi}^{\pi} \Delta x(e^{j\omega},\theta)^\top J \Phi_{\kappa}(\omega)J^*\Delta x(e^{-j\omega},\theta)d\omega = 0.
\eeq
The standard reasoning for showing uniqueness of the identification result is to show that if $\Eb \left[{||\Delta X(\theta)\xi_m||}_W\right]$ equals $0$ (i.e. when the minimum power is achieved), this should imply that $\Delta \bar G(\theta) = 0$ and $\Delta \bar H(\theta) = 0$. Since $J$ is full rank and positive definite, the above mentioned implication will be fulfilled only if
$\Phi_{\kappa}(\omega) > 0 $ for a sufficiently high number of frequencies. On condition 2 of Theorem \ref{theorem1} being satisfied along with the other conditions in Theorem 1, it ensures that the minimum value is achieved only for $\bar G(\theta) = \bar G^0$ and $\bar H(\theta) = \bar H^0$. \hfill $\Box$

\section{Proof of Proposition \ref{propx2}}
The disturbances in the original network are characterized by $\breve v$ (\ref{eq13xab}). From the results of Lemma \ref{lemx12}, we can infer that the spectral density $\Phi_{\breve{v}}$ has the unique spectral factorization
$\Phi_{\breve v} = \tilde H\Lambda\tilde H^*$ where $\tilde H$ is monic, stable, minimum phase, and of the form given in \eqref{eq20}. Together with the form of $\Lambda$ in \eqref{eq20} it follows that $\xi_{\smA}$ is uncorrelated with $\xi_{\smY}$. As a result, the set $\mA$ satisfies the properties of $\mF_n$, so that in Condition c we can replace $\mF$ by $\mB$. What remains to be shown is that the delay in path/loop conditions in the transformed network (\ref{eq5}) can be reformulated into the same conditions on the original network (\ref{eq1}). To this end we will need two Lemma's.
\begin{lemma}
	\label{lemma8}
	Consider a dynamic network as dealt with in Theorem \ref{theorem1}, with reference to eq. (\ref{eq5}), where a selection of node signals is decomposed into sets $\mD = \mQ \cup \mU$, $\mY = \mQ \cup \{ o\}$, and which is obtained after immersion of nodes in $\mZ$. Let $i$ be any element $i \in \mY\cup\mU$, and let $k$ be any element $k \in \mY$. \\
	If in the original network the direct path, as well as all paths that pass through non-measured nodes only, from $w_i$ to $w_{k}$ have a delay, then
	$\bar G_{ki}$ is strictly proper.
\end{lemma}
{\bf Proof:}
We will show that $\bar G_{ki}$ is strictly proper if all paths from $w_i$ to $w_k$ have a delay. For any $k \in \mY$, $i\in \mD$, $\bar G_{ki}$ is given by either (\ref{invarji1}) or (\ref{invarji2}) with $j=k$.
The situation that is not covered by (\ref{invarji1}), (\ref{invarji2}) is the case where $i=\{o\}$, but from (\ref{eq194xab}) it follows that $\bar G_{ko} = 0$, for $k \in \mY$. So for this situation strictly properness is guaranteed. \\
We will now use (\ref{invarji1}) and (\ref{invarji2}) for $j$ given by any $k \in \mY$.
In (\ref{invarji1}) and (\ref{invarji2}), it will hold that $\check H_{k3}$ is given by the appropriate component of (\ref{Hcheck}), which, by the fact that (\ref{eqspec}) is monic, will imply that $\check H_{k3}$ is strictly proper. By the same reasoning this also holds for $\check H_{ki}$.\\
From (\ref{invarji1}) and (\ref{invarji2}) it then follows that strictly properness of $\bar G_{ki}$ follows from strictly properness of $\breve G_{ki}$ if the inverse expression $(I-\breve G_{kk}+ \check H_{k3}\breve G_{\smU k})^{-1}$ is proper. This latter condition is guaranteed by the fact that $\check H_{k3}$ is strictly proper and $\breve G_{kk}$ and $(I-\breve G_{kk})^{-1}$ are proper as they reflect a module and network transfer function in the immersed network \cite{Woodbury&Dankers&Warnick_CDC:18,Weerts&etal_Autom:19_abstraction}.
Finally, strictly properness of $\breve G_{ki}$ follows from strictly properness of $G_{ki}$ and the presence of a delay in all paths from $w_i$ to $w_k$ that pass through unmeasured nodes.

%
\begin{lemma}
	\label{lemma9}
	Consider the transformed network and let $j, k$ be any elements $j,k \in \mY\cup\mU$. If in the original network all paths from $w_{k}$ to $w_j$
	have a delay, then all paths from $w_{k}$ to $w_j$ in the transformed network have a delay.
\end{lemma}
{\bf Proof:} This is proved using the Lemma 3 in \cite{VandenHof&etal_Autom:13} and Lemma \ref{lemma8}. Let $\bar G(\infty)$ denote $\lim_{z \rightarrow \infty} \bar G(z)$. From Lemma \ref{lemma8} we know $\bar{G}_{jk}$ is strictly proper if all paths from $w_k$ to $w_{j}$
in the original network have a delay. Therefore,
\beq
\bar G_m(\infty) = \begin{bmatrix}
	\ast & 0 \\ \ast & \ast
\end{bmatrix},
\eeq
where the 0 represents $\bar G_{jk}(\infty)$. Using inverse rule of block matrices we have,
\beq \label{eq56}
(I - \bar G_m(\infty))^{-1} = \begin{bmatrix}
	\ast & 0 \\ \ast & \ast
\end{bmatrix}
\eeq
Considering \eqref{eq5a} we can write $w_{m} = \bar G_m w_m + v_m$ where $v_m = \bar H_m\xi_m$. So have $w_{m} = (I - \bar G_m)^{-1}v_m$ where $(I - \bar G_m)^{-1}$ represents the transfer from $v_m$ to $w_m$. Having 0 in \eqref{eq56} represents that the transfer function from $v_k$ to $w_j$ has a delay. Since $v_k$ has path only to $w_k$ with unit transfer function, $w_k$ to $w_j$ has a delay.
  \hfill $\Box$

We now look into the proof of Proposition \ref{propx2}. For this we need to generalize the result  we have achieved in Lemma \ref{lemma9} in terms of scalar node signals to set of node signals. If all existing paths/loops from $w_{\smY\cup\smF}$ to $w_{\smY}$ in the original network have at least a delay, then all existing paths/loops from $w_k, k \in \mY\cup\mF$ to $w_j, j \in \mY$ in the original network have at least a delay. If all existing paths/loops from $w_k, k \in \mY\cup\mF$ to $w_j, j \in \mY$ in the original network have at least a delay, then as a result of Lemma \ref{lemma9}, all existing paths/loops from $w_k, k \in \mY\cup\mF$ to $w_j, j \in \mY$ in the transformed network have at least a delay. This implies that all existing paths/loops from $w_k, k \in \mY\cup\mF$ to $w_j, j \in \mY$ in the transformed network have at least a delay. Following the above reasoning, we can also show that if all existing paths from $w_{\smY\cup\smF}$ to $w_{k}, k \in \mF_n$ in the original network have at least a delay, all existing paths from $w_{\smY\cup\smF}$ to $w_k, k \in \mF_n$ in the transformed network have at least a delay.

\ifCLASSOPTIONcaptionsoff
  \newpage
\fi

\section*{Acknowledgment}
The authors gratefully acknowledge discussions with and contributions from Arne Dankers, Giulio Bottegal and Harm Weerts on the initial research that led to the presented results.

\bibliographystyle{IEEEtran}
\bibliography{Paul_Dynamic_Networks_Library}
%
\begin{IEEEbiography}[{\includegraphics[width=1in,height=1.25in,clip,keepaspectratio]{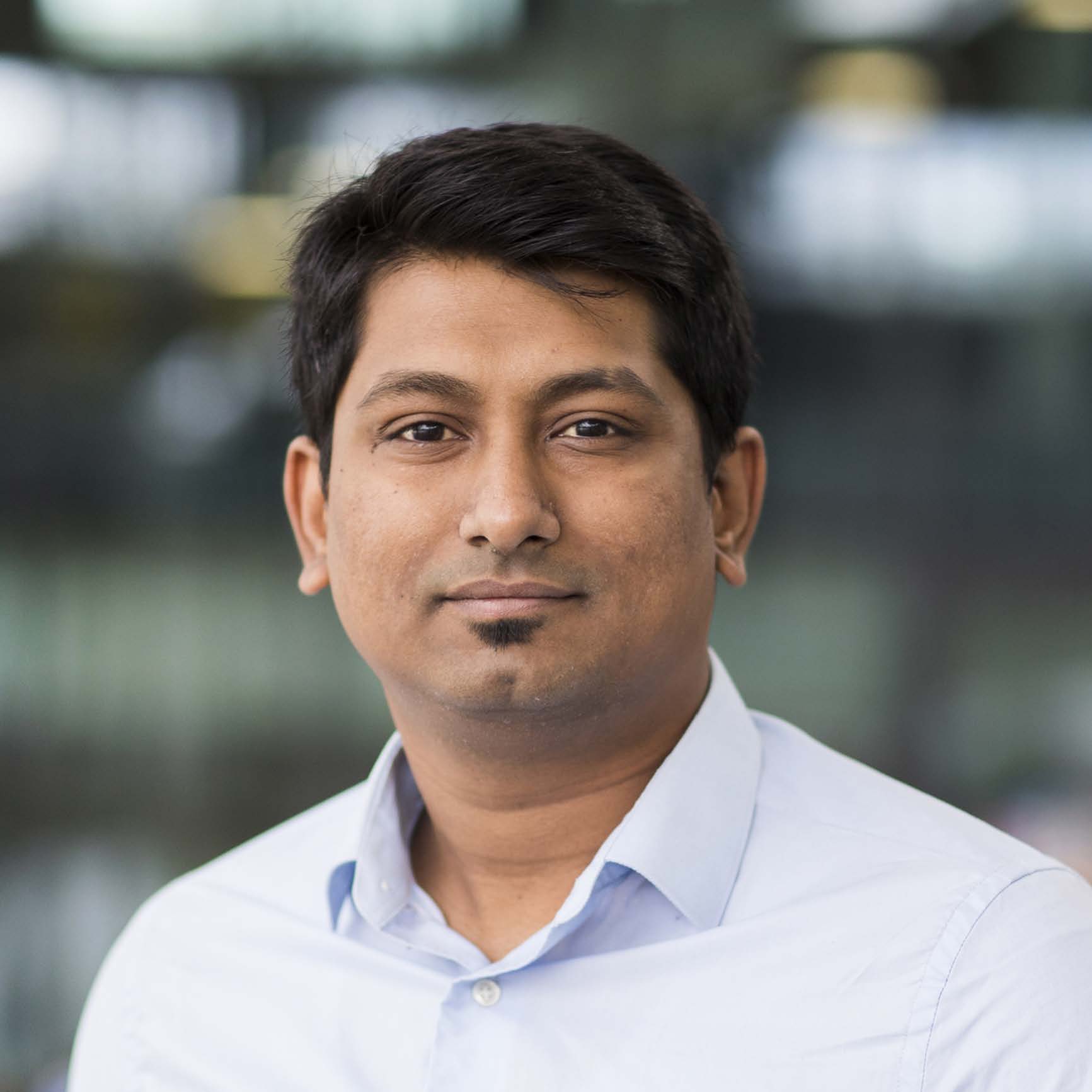}}]{Karthik Raghavan Ramaswamy} was born in 1989. He received his Bachelor’s in Electrical and Electronics Engineering (with Distinction) in 2011 from Anna University and Master’s in Systems and Control (with great appreciation) from TU Eindhoven in 2017. From 2011 to 2015 he was Control \& Automation engineer at Larsen \& Toubro. Currently, he is a PhD researcher with the Control Systems research group, Department of Electrical Engineering, TU Eindhoven, The Netherlands. His research interests are in the area of data driven modeling, dynamic network identification and machine learning.
\end{IEEEbiography}
\begin{IEEEbiography}[{\includegraphics[width=1in,height=1.25in,clip,keepaspectratio]{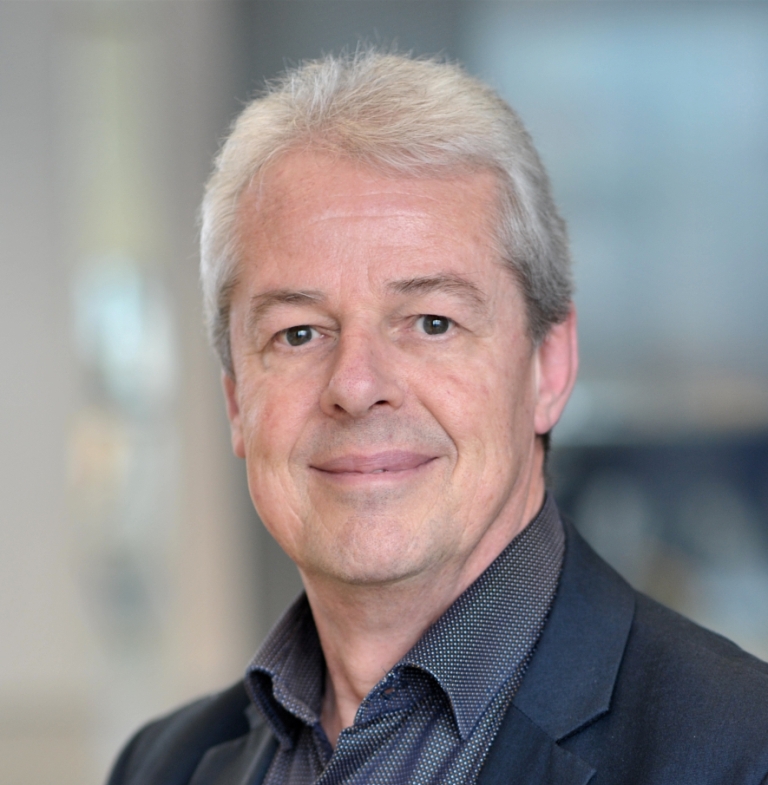}}]{Paul Van den Hof}
received the M.Sc. and Ph.D. degrees in electrical engineering from Eindhoven University of Technology, Eindhoven, The Netherlands, in 1982 and 1989, respectively.
In 1986 he moved to Delft University of Technology, where he was appointed as Full Professor in 1999. From 2003 to 2011, he was founding co-director of the Delft Center for Systems and
Control (DCSC). As of 2011, he is a Full Professor in the Electrical Engineering Department, Eindhoven University of Technology.
His research interests include data-driven modeling, identification for control, dynamic network identification,
and model-based control and optimization, with applications in industrial process control systems and high-tech systems. He holds an ERC Advanced Research grant for a research project on identification in dynamic networks.
Paul Van den Hof is an IFAC Fellow and IEEE Fellow, and Honorary Member of the Hungarian Academy of Sciences, and IFAC Advisor. He has been a member of the IFAC Council (1999–2005, 2017-2020), the Board of Governors of IEEE Control Systems Society (2003–2005), and an Associate Editor and Editor of Automatica (1992–2005). In the triennium 2017-2020 he was Vice-President of IFAC.
\end{IEEEbiography}
\end{document}